\documentclass{aa}
\usepackage[varg]{txfonts}

\usepackage{graphicx}
\usepackage{dcolumn}
\usepackage{bm}
\usepackage{wasysym}
\usepackage{amsmath}
\usepackage{amssymb}

\usepackage{euscript}

\newcommand{\fac}{f_\mathrm{acc}}

\begin{document}

\titlerunning{The central cusps in dark matter halos: fact or fiction?}
\title{The central cusps in dark matter halos: fact or fiction?}

\author{A.N. Baushev\inst{1}\and S.V.~Pilipenko\inst{2}}
\institute{Bogoliubov Laboratory of Theoretical Physics, Joint Institute for Nuclear Research,
141980 Dubna, Moscow Region, Russia \and Astro Space Center of P.N.Lebedev Physical Institute,
Russian Academy of Sciences, Profsoyuznaja 84/32, 117997 Moscow, Russia}
 \offprints{baushev@gmail.com}

\date{}

\abstract{We investigate the reliability of standard N-body simulations by modelling of the
well-known Hernquist halo with the help of \texttt{GADGET-2} code (which uses the tree algorithm to
calculate the gravitational force) and \texttt{ph4} code (which uses the direct summation).
Comparing the results, we find that the core formation in the halo center (which is conventionally
considered as the first sign of numerical effects, to be specific, of the collisional relaxation)
has nothing to do with the collisional relaxation, being defined by the properties of the tree
algorithm. This result casts doubts on the universally adopted criteria of the simulation
reliability in the halo center. Though we use a halo model, which is theoretically proved to be
stationary and stable, a sort of numerical 'violent relaxation' occurs. Its properties suggest that
this effect is highly likely responsible for the central cusp formation in cosmological modelling
of the large-scale structure, and then the 'core-cusp problem' is no more than a technical problem
of N-body simulations.}


\maketitle

\section{Introduction}
N-body simulations of the origin of present-day cosmic objects from initial small perturbations is
not only popular, but almost inevitable method of studying the large-scale structure formation in
the Universe: the task is highly nonlinear and complex, which makes numerical methods to be the
only direct approach to the problem. Simulating the structure formation under various cosmological
assumptions and comparing the results with observations one may check the applicability of the
cosmological models. Probably, the strongest result that has been derived from N-body cosmological
simulations is the core-cusp problem (see \citep{corecuspreview} for a review). Simulations of the
$\Lambda$CDM Universe always predict that dark matter (hereafter DM) halos have either a high
density core or a very steep density profile in the center \citep{gao2008, dutton2014} (for the
sake of simplicity, this dense central region is named 'a cusp', despite of its shape). However,
abundant observations of galaxies favor a respectively large and low-density cores in their central
regions: the dark matter mass there is much lower than the cusp should contain \citep{mamon2011,
walker2011}. This discrepancy is known for many years at the galaxy scale, but some recent
observations disfavor cusps in the galaxy clusters as well \citep{clustercores}.

The core-cusp problem may indicate that the cold dark matter model is incorrect. However, only
irreproachably reliable estimations of the simulation accuracy and convergence may permit to make
so strong physical conclusions. N-body simulations may suffer from several numerical effects; we
mention only the major ones.

Collisional relaxation of the test particles is an unphysical effect, if one models dark matter,
which is believed to be collisionless \citep{binney2002}. In the case of a spherical halo, the
process may be characterized by the relaxation time \citep[eqn. 1.32]{bt}
\begin{equation}
\tau_r =\dfrac{N(r)}{8\ln\Lambda}\cdot\tau_d \label{21a1}
\end{equation}
where $\ln\Lambda =\ln (b_{max}/b_{min})$ is the Coulomb logarithm ($b_{max}$ and $b_{min}$ are the
characteristic maximum and minimum values of the impact parameter), $N(r)$ is the number of test
particles inside radius $r$, $\tau_d=(4\pi G\bar\rho(r)/3)^{-1/2}$ is the characteristic dynamical
time of the system at radius $r$, $\bar\rho(r)$ is the average density inside $r$. It is
noteworthily that the potential softening used in the N-body simulations allows to avoid close
collisions of the particles, but scarcely affects the collisional relaxation time: $b_{min}$ is of
the order of the softening radius, but $\tau_r$ depends on it only logarithmically. We will use
$\ln\Lambda =\ln (3 r_s/\epsilon)$, where $\epsilon$ is the softening radius, like in \citep{17}.

The softening of the gravitational potential is an another source of computation errors. If
$\epsilon$ is too small, close two-body collisions on large angles occur. If the value of
$\epsilon$ is too large, it smooths away the high density peaks and introduces a bias in the force
computation. The optimal choice of softening has been a subject of many studies, see, e.g.
\citep{merritt96,2000MNRAS.314..475A,knebe2000,dehnen01,zhan06}.

The third potential source of undesirable numerical effects is the gravitational interaction
computations. The direct summation is exact, but very resource-intensive algorithm, since it
requires to calculate $N-1$ partial forces per particle in a system containing $N$ particles. It
means that we need $\sim N^2$ partial force calculations for all the system. Many N-body codes
 use a hierarchical tree algorithm \citep{tree1,tree2}, which is significantly more economical. The idea
of the method is to group distant particles into larger cells and calculate their gravity as a
single multipole force. A recursive subdivision of space is used to achieve the required spatial
adaptivity. For instance, the algorithm of the \texttt{GADGET} code \citep{springel2005} divides
the space in cubic cells. Each cell is repeatedly subdivided into eight daughter ones until the
ratio between the distance to the cell and the size of the cell exceeds the parameter specified by
the user. The \texttt{GADGET} code that we consider in this paper splits the cells until the
following cell-opening criterion is satisfied
\begin{equation}
\label{21b1} \frac{G M_{cell}}{r^2}\left(\frac{l}{r}\right)^2 \le \fac |\vec a|,
\end{equation}
where $M_{cell}$ and $l$ are the cell mass and extension, $r$ is the distance to the cell, $|\vec
a|$ is the value of the total acceleration obtained in the last timestep, and $\fac$ is a tolerance
parameter (see \citep[equation (18)]{springel2005}, where $\fac$ is denoted by $\alpha$). The
interaction with the closest particles is calculated by the direct summation. Requiring only
$O(N\ln N)$ partial force calculations for all the system, the tree algorithm is, generally
speaking, much faster than the direct summation. The price to pay is that the result represents
only an approximation to the true force.

Finally, figure 4 in \citep{springel2005} gives a highly visual illustration of significant
numerical phenomena appearing during the temporal integration of particle orbits. Thus, several
unphysical effects inevitably occur in the simulations, and their correct estimation is necessary
for an appropriate interpretation of simulation results.

However, the present state-of-art of N-body simulation tests (especially, of the DM behavior) can
hardly be named adequate. The commonly-used criterion of the convergence of N-body simulations in
the halo center is solely the density profile stability \citep{power2003}: these simulations show
that the central cusp (close to $\rho\propto r^{-1}$) is formed quite rapidly ($t<\tau_r$). The
shape of the cusp and its stability is insensitive to the simulation parameters (see, e.g.
\citep{hayashi2003}). Considering the temporal dependence of the overdensity inside various radii
$r$ from the halo center (i.e., the ratio of the average density inside radius $r$ from the halo
center to the average universe density), \citep{power2003} find that a core forms in the center no
earlier, than $t(r)= 1.7 \tau_r(r)$. We will denote this time by $\tau_p(r)= 1.7 \tau_r(r)$. The
authors assume that cusp universality and stability implies the negligibility of numerical effects,
and the core formation is the first sign of the collisional relaxation. Therefore,
\citep{power2003} suggest that N-body simulation results are trustable until $t(r)= 1.7 \tau_r(r)$.
The reasons why the collision influence can be neglected at almost two relaxation times are not
quite clear. Moreover, \citep{hayashi2003} and \citep{klypin2013} report that the cusp is stable
much longer, probably, up to tens of relaxation times.

The weak point of this reasoning is that the profile stability by itself does not guarantee the
absence of the collisional influence. Indeed, if the collisions are already significant, but the
particles mainly scatter on small angles in the collisions (the latter is true for N-body
simulations), the system can be described by the Fokker-Planck equation. This equation may have a
stationary solution \citep{evans1997, 13, 18}, and if it is stable, it works as a sort of
attractor: the collisions tend to transform an initial distribution into the stationary one. Then
the attractor profile should survive for much longer than $\tau_r$, since the Fokker-Planck
diffusion is self-compensated in this case. The profile corresponding to the stationary solution
should also be quite universal, since its shape is defined by the the Fokker-Planck coefficients
(i.e., by the potential of the particle interaction) that are similar for various N-body packages.
However, the profile universality and stability have nothing to do with the simulation veracity: it
is already created by test particle collisions, that is, by a purely numerical effect. It seems to
be no coincidence that the Fokker-Planck equation really has a stationary solution that is similar
to $\rho\propto r^{-1}$ in the halo center \citep{evans1997, 13} and close to the Einasto profile
with $n\sim 6$ at $r\sim 10^{-1} r_s$ \citep{18}.

Another way to test N-body simulations is to model an analytical solution and compare the results
with theoretical predictions. Simulations of an isolated spherically symmetric halo allow to go
beyond the density profile stability and consider the full array of dynamical parameters of the
particles. \citep{17} modelled a Hernquist halo and found that all integrals of motion
characterizing individual particles experience strong unphysical variations along the whole halo,
revealing an effective interaction between the test bodies. Moreover, the simulations show that the
cusp stability is really provided with the particle collisions, as we described in the previous
paragraph: intense upward and downward Fokker-Planck streams of particles in the cusp region occur,
and the cusp is stable because they compensate each other. This result suggests that the cusps in
cosmological N-body simulations may also be a consequence of numerical effects.

However, the paper \citep{17} has not answered to several important questions. Though significant
unphysical effects were found, the immediate causes of them, as well as their dependence from the
N-body code parameters, remained unclarified. The fact that the variations of the integrals of
motion of individual particles were significant at very large radii, where the influence of the
collisions and potential softening was certainly negligible, implied that the integral variations
there are most likely due to the potential calculating algorithm. In order to clarify this point,
we perform a new simulation of the Hernquist halo, following the way described in \citep{17}.
However, contrary to that work, we evolve the same initial conditions for the halo using the
\texttt{GADGET-2} code with various parameter settings. Besides, we follow the system evolution in
a 4-th order Hermite code, which uses the direct summation algorithm to calculate the gravitation
force, and thus is free from the tree algorithm drawbacks. Comparing the results, we may elucidate
the cause of the unphysical effects and clarify their dependence from the code parameters.

In Sect.~\ref{21sec2} we describe the codes we used and the simulation setup, in Sect.~\ref{21sec3}
we present the methods we use to treat the data, in Sect.~\ref{21sec4}, we present the results of
the simulations and discuss them. Finally, in Sect.~\ref{21sec5} we briefly summarize the obtained
results.

\section{Simulations}
\label{21sec2}

\subsection{The simulation setup}
We evolve the system using one of the most extensively employed in cosmological simulations SPH
codes, \texttt{GADGET-2} \citep{springel2001, springel2005}\footnote{Actually, the dark matter part
of the \texttt{GADGET-3} code coincides with that of the \texttt{GADGET-2}. The codes differ only
in the barionic matter part.}. We vary the cell-opening parameter $\fac$ to check the impact of the
tree algorithm on the system evaluation. The lower the value of $f_\mathrm{acc}$ is, the more
\texttt{GADGET-2} resembles a direct summation algorithm with the second order accuracy of temporal
integration. The default value of $\fac$ recommended by \citep{springel2005} is $5\cdot 10^{-3}$.
So we use three values: $\fac=5\cdot 10^{-5}$, $5\cdot 10^{-4}$, and $5\cdot 10^{-3}$. For
comparison, we also use a 4-th order direct summation parallel Hermite code with individual
particle time steps, which is a part of \texttt{AMUSE} suite \citep{amuse1,amuse2} and is called
``\texttt{ph4}'' inside it. Of course, the direct summation code is much slower than the tree one.
All \texttt{ph4} simulations were run at the Lomonosov supercomputer of the Moscow State University
computer center\footnote{http://parallel.ru}.

Thus, the \texttt{GADGET-2} and \texttt{ph4} codes have significantly different field calculation
algorithms. However, they differ little in other calculation performance: both the codes utilize
the leapfrog algorithm to calculate the test particle trajectories, and very similar gravitational
softening to avoid close particle collisions. In \texttt{ph4}, the value of the softening radius
$\epsilon$ gives the radius of a Plummer `sphere', i.e. the gravitational potential of a point mass
is calculated as $-Gm/\sqrt{r^2+\epsilon^2}$. The \texttt{GADGET-2} code uses the same value of the
potential at zero radius $-Gm/\epsilon$ and  represents the density distribution of a point mass by
a spline kernel. Thus, despite of some difference in gravitational softening, the meaning of the
softening parameter is the same for both codes. We ignore all the data inside $0.1 a=10$~{pc}
(i.e., $\sim 6$ softening radii $\epsilon$ if $\epsilon = 1.7$~{pc} and $\sim 3 \epsilon$ if
$\epsilon = 3.4$~{pc}) from the halo center in order to avoid the influence of the potential
softening on the cusp.

\begin{figure}
 \resizebox{\hsize}{!}{\includegraphics[angle=0]{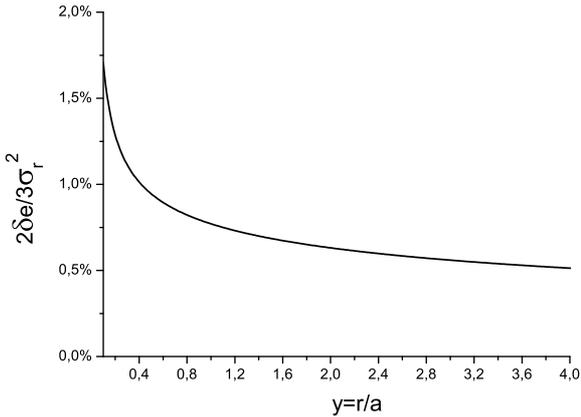}}
\caption{The ratio of the perturbation energy of the clump particles $\delta e$ (see
eqn.~\ref{21e3}) to the average kinetic energy of the particles $\bar e_k\simeq 3 \sigma^2_r/2$, as
a function of radius.}
 \label{21poisson}
\end{figure}

\subsection{The Hernquist profile}
The Hernquist profile has the shape
\begin{equation}
\label{21b2} \rho_{ini}(r)=\frac{Ma}{2\pi r(r+a)^3},\qquad M_r(r)=M\frac{r^2}{(r+a)^2}
\end{equation}
where $M$ is the total halo mass, $M_r(r)$ is the halo mass inside radius $r$, $a$ is the Hernquist
radius, $r_s=a/2$ \citep{hernquist1990}. It is important that it behaves exactly as the NFW in the
center. The specific angular momentum $K_{circ}$ corresponding to a circular orbit of radius $r$ is
\begin{equation}
\label{21b3} K_{circ}(r)=\sqrt{G M}\frac{r^{3/2}}{(r+a)}.
\end{equation}
\noindent We set $M = 10^9 M_\odot$, $a=100$~{pc}, which roughly corresponds to the well-known
dwarf satellite of the Milky Way, Draco, as it was before the tidal destruction \citep{draco2002}.
Since we use the standard N-body units \citep{nbodyunits}, the results are independent on the
choice of $a$ and halo mass. However, it defines the time scale, and we choose the realistic values
for illustrative purposes.

It is convenient to introduce the Hernquist time $\xi=\left(\dfrac{GM}{a^3}\right)^{-1/2}$ and the
dimensionless radius $y\equiv r/a$. Then
\begin{equation}
\tau_d= \xi (y+1) \sqrt{y};\qquad \tau_r=\dfrac{N\xi}{8 \ln\Lambda}\dfrac{y^2\sqrt{y}}{(y+1)}.
 \label{ap21a4}
\end{equation}
Here $N$ is the total number of particles in the Hernquist halo.

In order to generate the initial conditions, we place randomly $N\sim 10^5$ test bodies of equal
masses, in accordance with the analytically obtained space and velocity distributions,
corresponding to the Hernquist profile \citep{hernquist1990}. Theoretically, the halo should be
stable: not only the density profile, but even the energy and angular momentum components of each
particle should be conserved. Any deviation from this behavior is a numerical effect. The
relatively small number $N$ of test particles comes from the slowness of the direct summation
algorithm: simulations for $N=10^6$ particles would last for too long. On the other hand, even
recent and high-performance cosmological simulations contain only a few of halos with $\sim 10^6$
test particles \citep{dutton2014}.

Our main simulation contains $N=10^5$ particles. We choose the softening length $\epsilon = 0.5 a
N_a^{-1/3}$, where $N_a$ is the number of particles inside radius $a$ around the halo center. This
value of the gravitational softening length ($\epsilon = 1.7$~{pc} in our case) is found to be
optimal for the Hernquist profile \citep{dehnen01}. Then the Coulomb logarithm is equal to
$\ln\Lambda \simeq 4.5$ for the halo under consideration, $\xi=4.72*10^{6}$~{years}, and

\begin{equation}
\tau_d= (y+1) \sqrt{y}\times 0.472\cdot 10^{6} \text{years};\qquad \tau_r=\dfrac{2
y^2\sqrt{y}}{(y+1)}\times 1.36\cdot 10^{9} \text{years}.
 \label{21a4}
\end{equation}
At $r=a$ $\tau_d=2\cdot\xi=0.945\cdot 10^{6}$~{years}, $\tau_r=1.36\cdot 10^{9}$~{years}. We
consider this simulation as the main one, and it covers the maximum duration $2.85\cdot
10^{9}\text{years}=2.85$~{Gyr}. Hereafter we always consider this simulation in these paper, unless
the opposite is indicated. If the name of a simulation is not specified somewhere in the text, we
mean this one.

To check the result reliability, we perform an auxiliary simulation with the same number of
particles ($10^5$), but with the doubled softening length $\epsilon = 3.4$~{pc}. Moreover, we
create the initial condition setting the random number generator independent from the first
simulation, which gives us a possibility to estimate the importance of the Poisson noise (see
below). Since we are short of calculational power, the simulation covers only $0.85\cdot
10^{9}\text{years}=0.85$~{Gyr}. Every time when we use this auxiliary simulation, we specify it, so
it may not lead to a misunderstanding.

We model an astrophysical stationary DM halo, which has a smooth and constant gravitational
potential $\phi(r)$. Therefore, the specific energy $w=\phi(r)+v^2/2$ and the specific angular
momentum $\vec K$ of each particle should be conserved\footnote{Sometimes we will name 'specific
energy' and 'specific angular momentum' simply by 'energy' and 'angular momentum'. Since we never
use real energies and angular momenta of the particles in this paper, it cannot lead to any
misunderstanding.}. Instead of $w$, it is more convenient to use the apocenter distance of the
particle $r_0$ (i.e. the maximum distance on which the particle can move off the center, which can
be found from the implicit equation $w=\phi(r_0)+K^2/2r_0$).  We analyze the behavior of $K$,
$K_x$, $K_y$, $K_z$, and $r_0$ ($r_0$ is an integral of motion as an implicit function of the
integrals of motion $w$ and $K$) of individual particles in our simulations. Any temporal variation
of these values is necessarily a numerical effect.

\begin{figure*}[p]
\resizebox{\hsize}{!}{\includegraphics[angle=0]{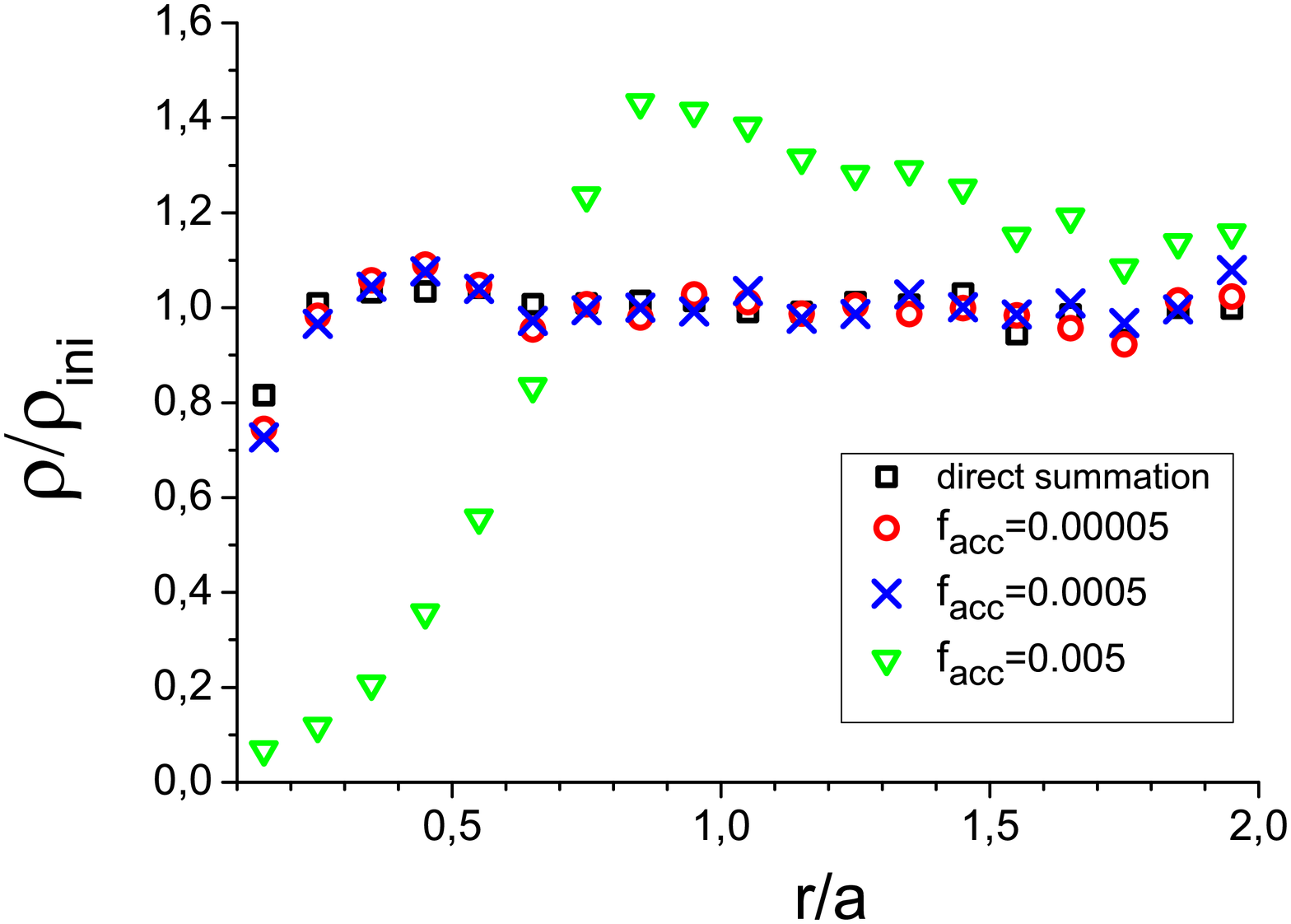}}
\resizebox{\hsize}{!}{\includegraphics[angle=0]{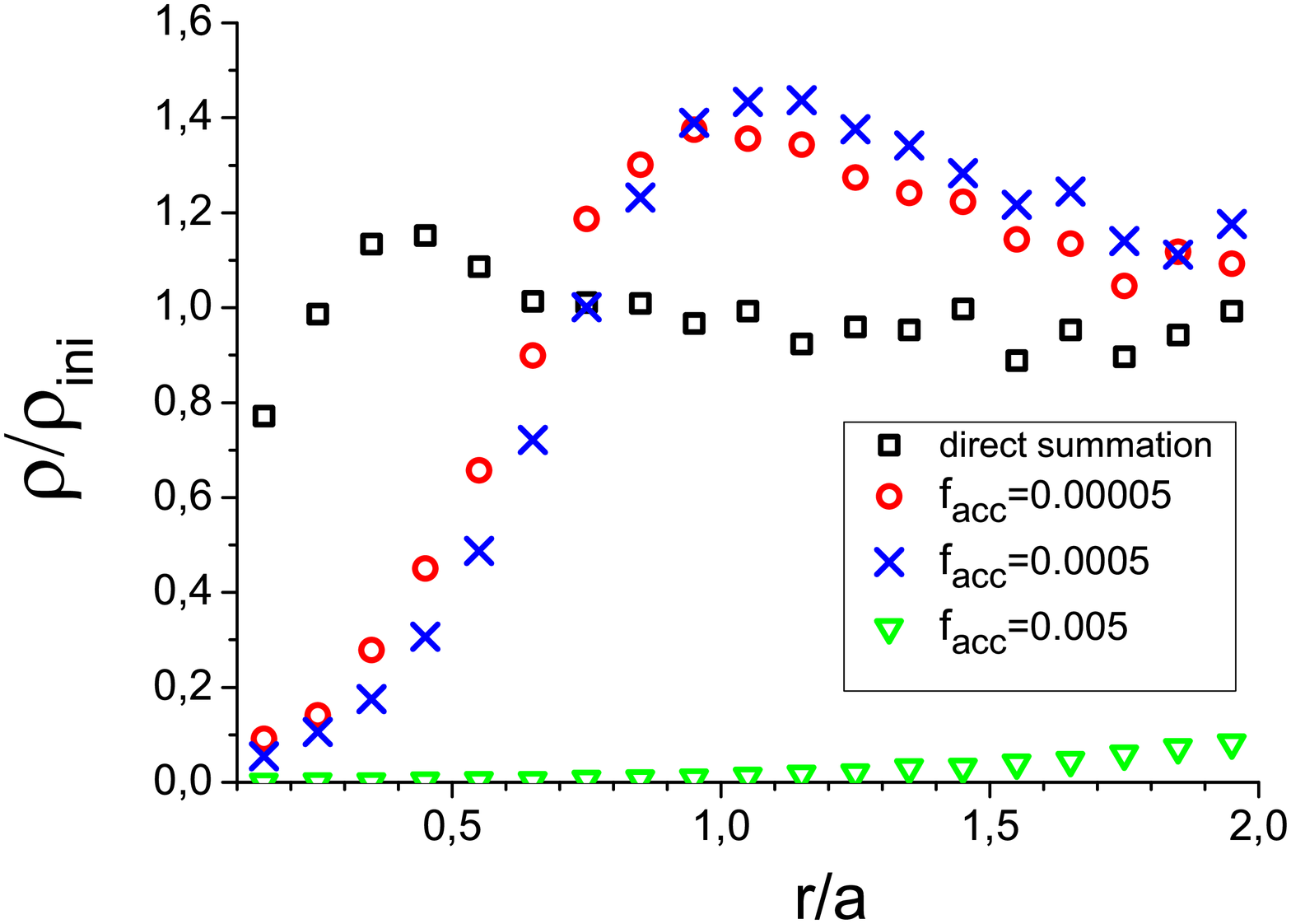}} \caption{The density profiles
at $t=0.45\cdot 10^9\text{years}=0.45$~{Gyr} (top panel) and $t=2.85\cdot
10^{9}\text{years}=2.85$~{Gyr} (bottom panel). Black squares, red circles, blue crosses, and green
triangles correspond to the direct summation (\texttt{ph4} code), $\fac=5\cdot 10^{-5}$, $5\cdot
10^{-4}$, and $5\cdot 10^{-3}$, respectively.}
 \label{21fig1}
\end{figure*}

\subsection{Is the Poisson noise significant?}
\label{poisson}

However, the numerical effects may be produced not only by N-body codes, but by inaccuracies of the
initial conditions as well. The only significant source of errors in the initial conditions we use
is the Poisson noise: since the particles are placed randomly, they may stochastically form clumps,
and we should estimate the influence of this effect. Let us consider a Poisson clump of mass
$\mathfrak{m}$ occurred at a radius $l$ (and dimensionless radius $y\equiv l/a$) from the halo
center. The number of particles in the clump is $\sim N\mathfrak{m}/M$, where $N$ and $M$ are the
total mass and the number of particles in the halo. The standard deviation of this quantity is
$\sim \sqrt{N\mathfrak{m}/M}$, and the standard deviation of the clump mass is
$\delta\mathfrak{m}\sim \mathfrak{m}\sqrt{M/(N\mathfrak{m})}=\sqrt{M\mathfrak{m}/N}$.

\noindent Two questions should be discussed:
\begin{enumerate}
\item Will the clump grow? \item What characteristic perturbations of the integrals of motion
$K_x$, $K_y$, $K_z$, and $r_0$ may the Poisson noise produce?
\end{enumerate}

\noindent The answer on the first one is certainly negative. First, let us suppose that the clump
of radius $r$ is gravitationally bound (we will see that it is not really so), but even then the
clump will be destroyed by the tidal force of the main halo. Indeed, the characteristic size of
many particle orbits in the clump of radius $r$ is $r$. The tidal gravitational force from the main
halo can be estimated as $\sim (G M(l)/l^2)\cdot (r/l)$ in the framework of the clump center. The
work of the force on the particle orbit is $\sim r (G M(l)/l^2)\cdot (r/l)$. If it exceeds the bind
energy of the particle $\sim G\mathfrak{m}/r$, the particle can be teared out of the clump: $r(G
M(l)/l^2)\cdot (r/l)>G\mathfrak{m}/r$, or $(M(l)/l^3)>(\mathfrak{m}/r^3)$. This inequality may be
rewritten as $\bar\rho(l)>\rho(l)$, where $\bar\rho(l)$ and $\rho(l)$ are the average halo density
inside radius $l$ and the halo density at radius $l$, respectively: the relative overdensity of the
clump is small $\delta\mathfrak{m}/\mathfrak{m}\sim 1\sqrt{N}$ and may be neglected. But the
inequity $\bar\rho(l)>\rho(l)$ is always true, since the central region of the halo is denser.
Thus, tidal perturbations exceed the binding energy of the Poisson clumps, and they are easily
destroyed. This is a special case of a general regularity: a clump can be stable only if its
density is significantly higher than the halo density \citep{19}.

Second, the Poisson clumps in our system cannot be gravitationally bound at all. Then $r\sim
\sqrt[3]{3\mathfrak{m}/(4\pi \rho(l))}$, where we should substitute (\ref{21b2}) instead of
$\rho(y)$.
\begin{equation}
r\sim \sqrt[3]{\frac{3\mathfrak{m}}{4\pi \rho(l)}},
 \label{21e1}
\end{equation}
where we should substitute (\ref{21b2}) instead of $\rho(l)$.   The standard deviation of particle
energy $\delta e$ produced by the clump is $\delta e= G \delta \mathfrak{m}/r$. We obtain:
\begin{equation}
\delta e\sim 0.8 \frac{GM}{a} N^{-1/2} \frac{(\mathfrak{m}/M)^{1/6}}{\sqrt[3]{y}(y+1)}.
 \label{21e2}
\end{equation}
As we can see, the Poisson noise depends on the clump mass rather weakly, only as
${\mathfrak{m}}^{1/6}$.

Deriving this equation, we supposed (equation~\ref{21e1}) that the average density of the clump is
equal to to the clump central density. This is apparently not true if $r\gg l$: then the clump
center is very close to the halo center, and $\rho(l)$ by far exceeds the average clump density. In
fact, the average clump density cannot exceed
 \begin{equation}
 \min\left(\sqrt[3]{\frac{3M(l)}{4\pi l^3}}, \sqrt[3]{\frac{3 M(r)}{4\pi r^3}}\right).
 \label{21g1}
 \end{equation}
As a result, equation~\ref{21e2} for $\delta e$ diverges as $\sqrt[3]{y}$ when $y\to 0$. In order
to avoid this difficulty, we limit ourself by considering only the case when the halo center does
not lie inside the clump, i.e., $r<l$. Then $\mathfrak{m}$ by no means can exceed $M(l)$. With the
help of equation~(\ref{21b2}) we obtain:
 \begin{equation}
 \frac{\mathfrak{m}}{M}<\frac{M(l)}{M}=\frac{y^2}{(y+1)^2},\quad \text{i.e.} \quad
 \frac{(\mathfrak{m}/M)^{1/6}}{\sqrt[3]{y}}<(y+1)^{1/3}
 \label{21g2}
 \end{equation}
With the help of this inequality, we obtain from~(\ref{21e2})
\begin{equation}
\delta e\sim 0.8 \frac{GM}{a} N^{-1/2}(y+1)^{-4/3}.
 \label{21e3}
\end{equation}
The characteristic velocity of the particles in the clump is the particle velocity dispersion
$\sigma$, corresponding to the Hernquist profile. Since the velocity distribution is isotropic, the
velocity dispersions are equal in all three directions, i,e. the average squared velocity of the
particles is $v^2\sim 3 \sigma^2_r$. The measure of the Poisson clump stability is the ratio
$\alpha\equiv 2\delta e/v^2=2\delta e/3 \sigma^2_r$. Calculation of $\sigma^2_r$ for the Hernquist
profile is trivial, but rather cumbersome \citep[equation 4.35b]{bt}.

Figure~\ref{21poisson} shows the ratio $\alpha= \delta e/\bar e_k=2\delta e/3 \sigma^2_r$ (i.e.,
the ratio of the gravitational binding energy of the clump particles to the average kinetic energy
of the particles) as a function of radius. As we can see, $\alpha=2\delta e/3 \sigma^2_r$ does not
exceed $2\%$ for $r\ge 0.1 a$, i.e., the average kinetic energy of the particles exceeds the
binding energy by more than $50$ times.

Thus, the clump grows (or even stability) is out of the question: the clump flies apart in a time
$\sim r/\sigma(l)\ll \tau_d(l)$, and its particles are randomly spread over the halo at radius
$\sim l$ in a time $\sim \tau_d(l)$. However, their traces remain in the phase space of the system
\citep{bt}: some particles are accelerated and some particles are decelerated by the gravitational
field of initial Poisson noise, and so their integrals of motion are perturbed with respect to the
exact analytical solution. Equation~\ref{21e3} allows us to estimate the integral variations via
this effect. Indeed, $\alpha\simeq\delta e/\sigma^2_r=\delta (v^2/2)/\sigma^2_r\simeq v\delta
v/\sigma^2_r\simeq \delta v/v$. It means that the ratio error of the particle speed occurring as a
result of the Poisson noise is $\sim 2\%$ at $r\sim 0.1 a$ and decreases with radius. The same
estimation is valid for the ratio errors of the integrals of motion.

Thus, the simulations of a formes halo are not that sensitive to the Poisson noise, as a result of
significant velocity dispersion in the formed halo, and stochastically generated initial conditions
may be quite acceptable in this case. Below we will revert to the influence of the Poisson noise in
our simulations.

\section{Data treatment}
\label{21sec3}

First of all, we considered the velocity distribution of the particles at various radii and
temporal moments and found no anisotropy: the initial isotropy of the velocity distribution
conserves. We ignore all the data inside $0.1 a=10$~{pc}, i.e., $\sim 6\epsilon$ (or $\sim
3\epsilon$ for the second simulation with $10^5$ particles) from the halo center in order to avoid
the influence of the potential softening on the cusp. Moreover, our statistics is poor in this
region.

In order to consider the density profile evolution, we split the halo into spherically symmetric
layers of the same thickness $10\text{pc}=0.1a$ and divide the number of particles in each by the
value corresponding to the initial Hernquist profile (\ref{21b2}). Thus we obtain an approximation
of the current halo density as a fraction of the initial density $\rho_{ini}(r)$ given by the
equation (\ref{21b2}).

We want to consider variations of the integrals of motion and their dependence of radius. We split
the $10^5$ particles in simulations into groups of $1600$ units in each, according to their initial
$r_0$, starting from the lowest $r_0$. Thus, all the particles in the same group have similar
$r_0$. In principle, the group may be characterized by the averaged initial $\overline{r_0}$ of its
members. Indeed, if the particle orbit is elongated, the particle spends almost all the time near
the apocenter, in accordance with the Kepler's second law. On the contrary, if the orbit is
circular, the particle moves along almost uniformly, but its radius always remains close to $r_0$.
However, each particle contributes to the density profile on an interval between its pericenter
radius $r_{min}$ and apocenter radius $r_0$. Therefore, we use the averaged initial radius of the
particles of the group $\overline{r_{ini}}$ as the characteristic radius corresponding to the
group. This choice is not very important: $\overline{r_{ini}}\lesssim \overline{r_0}$ for each
group.

We examine the behavior of five integrals of motion: $K$, $K_x$, $K_y$, $K_z$, and $r_0$. Let us
use $K$ to illustrate the procedure. At various given temporal moments, we calculate the change
$\Delta K$ of $K$ with respect to the initial value for each particle of a group. Then we evaluate
the mean $\mu(\Delta K)$ and the standard deviation $\sigma(\Delta K)$ over the group.

To visualize the significance of the variations of the integrals of motion, we transfer to
dimensionless values. We divide the mean changes $\mu$ and $\sigma$, relating to the angular
momenta and their components, by $K_{circ}(\overline{r_0})$, the angular momentum corresponding to
the circular orbit at $\overline{r_0}$. Roughly speaking, $K_{circ}(\overline{r_0})$ is the maximum
value of $K$ any particle with the apocenter distance $\overline{r_0}$ may possess. We divide the
mean changes $\mu (\Delta r_0)$ and $\sigma (\Delta r_0)$ by $\overline{r_0}$. Being so defined,
the values of $\mu$ and $\sigma$ give us the idea of the importance of the non-conservation of the
integrals of motion at various moments of time.

The non-conservation of integrals of motion leads to the Fokker-Planck diffusion of particles in
the phase space. Indeed, \citep{17} found strong Fokker-Planck streams in the cusp area, which
suggested that the numerical effect of alteration of the integrals of motion might significantly
influence the shape of the cusp or even create it.

In order to investigate the point, we follow the procedure offered in \citep{17}. For an array of
radii $r$, we calculate the number $N_{+}(t,r)$ of particles that have $r_0<r$ at the initial
moment and $r_0>r$ at the time $t$, and the number $N_{-}(t,r)$ of particles that have $r_0>r$ at
the initial moment and $r_0<r$ at the time $t$. By counting each particle no more than once, we
avoid a possible effect of small noise, produced by particles having $r_0$ just near the boundary
radius $r$, crossing and recrossing it and thus giving an impression of intensive flows that do not
exist. On the other hand, the values of $N_{+}(t,r)$ and $N_{-}(t,r)$ give only the lower bounds on
the upward and downward Fokker-Planck streams of particles: the value of $r_0$ of a particle could
have crossed $r$ an odd number of times (and then it is counted only once) or an even number of
times (and then it is not counted at all). In the collision-less case,  $N_{+}(t,r)$ and
$N_{-}(t,r)$ are obligatory equal to zero, since $r_0$ is an integral of motion.

\section{Results and Discussion}
\label{21sec4}

Before discussing the obtained results, it is pertinent to remind briefly the main possible sources
of numerical effects and the differences between the N-body codes we use. The discretization may
lead to numerical artefact occurrence, the collisional relaxation being one of them. All the
simulations we perform in this work are equivalent in this sense, since the initial conditions (in
particular, the number of particles in the halo) are exactly the same, and the potential softening
is identical (which means that the Coulomb logarithm $\ln\Lambda$ has the same value). N-body codes
typically use relatively simple algorithms of particle trajectory evaluation, for calculation
duration's sake, which may also lead to numerical effects. However, \texttt{GADGET-2} and
\texttt{ph4} use the same leapfrog algorithm. Thus, the only significant difference between the
four simulations of the same halo that we perform in this work is the potential calculation
algorithm. The most precise is the direct summation algorithm realized in \texttt{ph4}. The
accuracy of the tree algorithm of \texttt{GADGET-2} code depends on the cell-opening parameter
$\fac$: augmentation of $\fac$ makes the algorithm faster, but increases the numerical effects as
well. We use three values: $\fac=5\cdot 10^{-5}$, $5\cdot 10^{-4}$, and $5\cdot 10^{-3}$; the last
one is typical for cosmological N-body simulations.

\subsection{Density profiles and convergence criteria of N-body simulations}

Figure~(\ref{21fig1}) represents the density profiles (obtained with the help of the procedure
described in the previous section) at $t=0.45\cdot 10^9\text{years}=0.45$~{Gyr} and $2.85$~{Gyr}.
Black squares, red circles, blue crosses, and green triangles correspond to the direct summation
(\texttt{ph4} code), $\fac=5\cdot 10^{-5}$, $5\cdot 10^{-4}$, and $5\cdot 10^{-3}$, respectively.

We can see that the cusp behavior drastically depend on the potential calculation algorithm. The
top panel in figure~(\ref{21fig1}) corresponds to $\sim \tau_r$ at the radius $0.6 a$ and to the
Power's time $\tau_p=1.7 \tau_r$ at the radius $0.45 a$ (the relaxation and Power's times scale
with radius in accordance with (\ref{21a4})). The profiles calculated by the \texttt{ph4} code and
by \texttt{GADGET-2} with $\fac=5\cdot 10^{-5}$ and $\fac=5\cdot 10^{-4}$ are still very similar
and show yet no core, just a small drawdown in the center. A clear core with radius $r_{core}\simeq
0.6 a$ forms in the case of the \texttt{GADGET-2} code with $\fac=5\cdot 10^{-3}$.

The bottom panel in figure~(\ref{21fig1}) corresponds to $\sim \tau_r$ at the radius $1.45 a$ and
to the Power's time $\tau_p$ at the radius $1.1 a$. The core radius is $r_{core}\simeq 0.6 a$ for
$\fac=5\cdot 10^{-5}$ and $\fac=5\cdot 10^{-4}$, and exceeds $2a$ for $\fac=5\cdot 10^{-3}$. We
should make a remark that, though the central part of the density profile corresponding to $5\cdot
10^{-3}$ seems to be completely emptied at the bottom panel of figure~(\ref{21fig1}), the absolute
value of density still has a high peak in the center in this case. The impression of emptiness
occurs because the density drops very significantly with respect to the initial cuspy profile. The
most impressing fact is that the core does not form at all if we use the \texttt{ph4} code: $\sim
14 \tau_r$ have already passed at $r=0.4 a$, and there is no sign of core there. We can see only a
small central drawdown.

We may make several conclusions from the plots. First, while the \texttt{GADGET-2} profiles
corresponding to $\fac=5\cdot 10^{-5}$ and $\fac=5\cdot 10^{-4}$ and the \texttt{ph4} profile
branch off after $\sim 1$~{Gyr}, the behavior of the profiles for the \texttt{GADGET-2} simulations
with $\fac=5\cdot 10^{-5}$ and $\fac=5\cdot 10^{-4}$ is identical. On the one hand, this might be a
hint that a decreasing of $\fac$ below $5\cdot 10^{-4}$ is not effective: it significantly
increases the evaluation time and does not improve the result (at least, for the system under
consideration). On the other hand, there is a popular way to check the simulation convergence: vary
the simulation parameters by several times and (provided that the simulation results do not change)
consider it as a prove of the simulation reliability. A comparison with the results of the direct
summation algorithm shows that the criterion fails in our case: though the decreasing of $\fac$
from $5\cdot 10^{-4}$ to $5\cdot 10^{-5}$ does not change the simulation results, the direct
summation algorithm is still much better.

Second, the core formation occurs a bit earlier, but almost exactly as reported in
\citep{power2003} if $\fac=5\cdot 10^{-3}$. The core forms much later than the Power's time if
$\fac=5\cdot 10^{-4}$ or $5\cdot 10^{-5}$. This is not surprising: $\fac=10^{-3}$ was used by
\citep{power2003}.

The third and the main conclusion is that (contrary to contrary to popular belief) the core
formation in the halo centers in cosmological N-body simulations has nothing to do with the
collisional relaxation of particles, being totaly defined by the characteristics of the evaluation
algorithm of gravitational field. Indeed, the collisional relaxation should manifest itself in
exactly the same way for all the values of $\fac$: we have exactly the same number and masses of
particles, and exactly the same initial distribution. The softening radius $\epsilon$ is also the
same, and so is the Coulomb logarithm $\ln\Lambda$. Nevertheless, the profile behaviors are
completely different, i.e., they are defined by the value of the tree algorithm parameter $\fac$,
and not by the collisional relaxation.

This illustrates the failure of the generally recognized criteria \citep{power2003} of N-body
simulation reliability, based on the profile stability and universality. Indeed, the cusp is
routinely formed in the halo centers, being quite insensitive to the numerical code realization and
initial conditions, then it remains stable until $t\sim 1.7 \tau_r(r)$, and then a core appears. It
is widely believed that the cusp stability and universality proves the negligibility of numerical
effects until $t\sim 1.7 \tau_r(r)$, while the core formation is the first manifestation of
numerical effects (namely, of the collisional relaxation).

However, as we could see, the start of the core formation has nothing to do with the collisional
relaxation; it is mainly defined by the properties of the tree algorithm, and can be strongly
postponed or even avoided by the use of the direct summation algorithm \texttt{ph4}. As for the
numerical effects, we will see in the next subsection that they become quite tangible on the time
scale $\sim \tau_d(r)$, i.e., orders of magnitude earlier than the core formation. Thus, the
commonly-accepted use of the central core formation as the first sign of presence of numerical
effects breaks down. As we will see, the cusp formation itself is probably due to numerical
effects.

\subsection{Numerical 'violent relaxation'}

\begin{figure*}[p]

\begin{tabular}{p{\textwidth}}

 \resizebox{0.48\hsize}{!}{\includegraphics[angle=0]{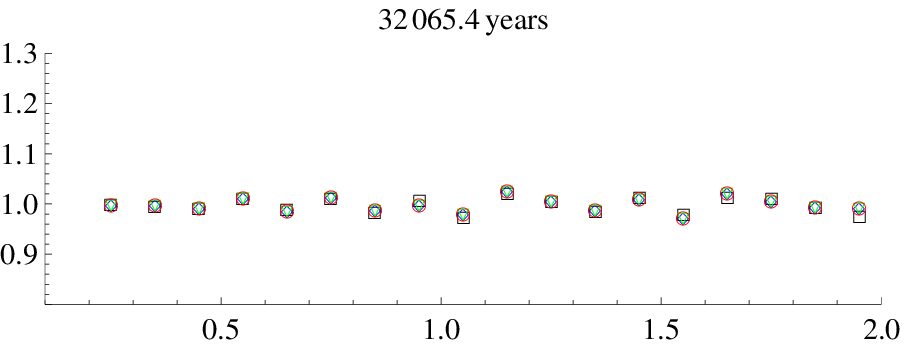}}
 \resizebox{0.48\hsize}{!}{\includegraphics[angle=0]{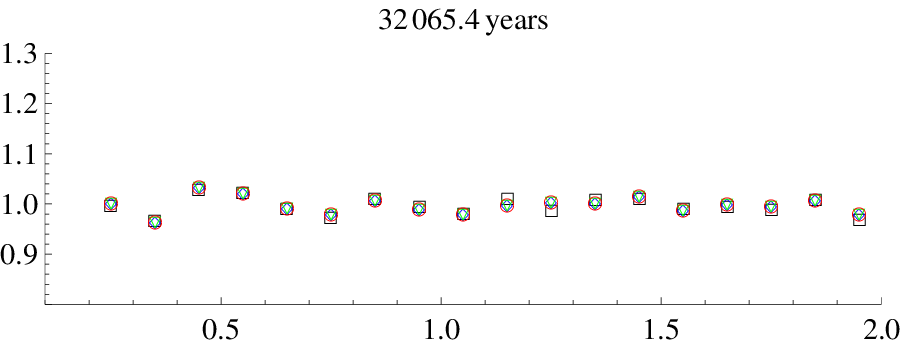}}\\

 \resizebox{0.48\hsize}{!}{\includegraphics[angle=0]{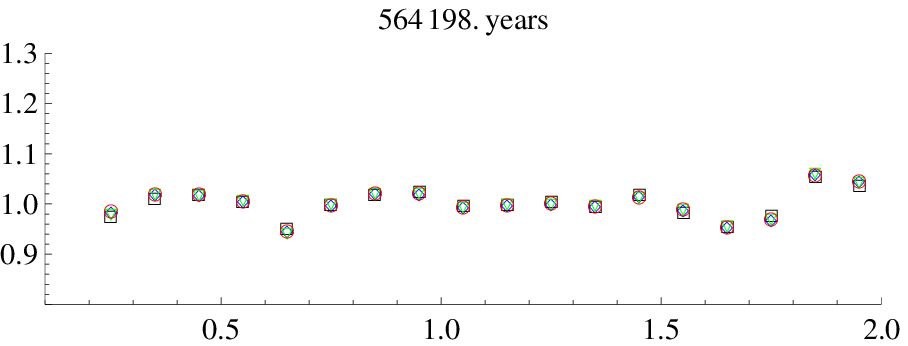}}
 \resizebox{0.48\hsize}{!}{\includegraphics[angle=0]{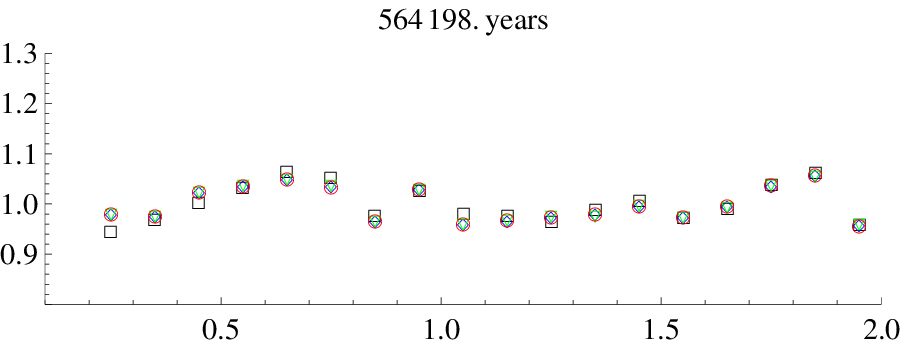}}\\

 \resizebox{0.48\hsize}{!}{\includegraphics[angle=0]{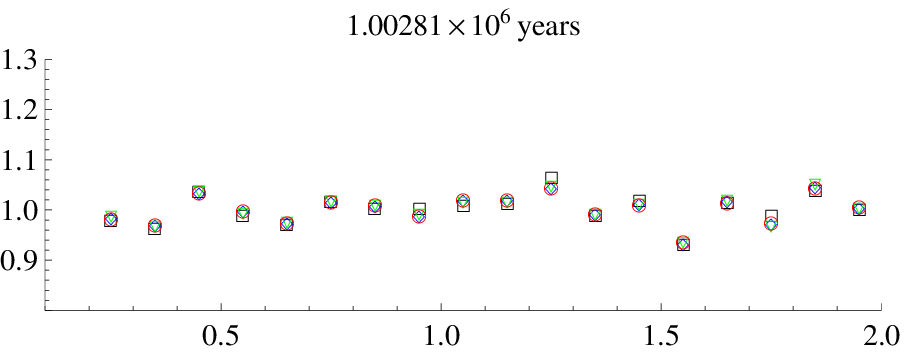}}
 \resizebox{0.48\hsize}{!}{\includegraphics[angle=0]{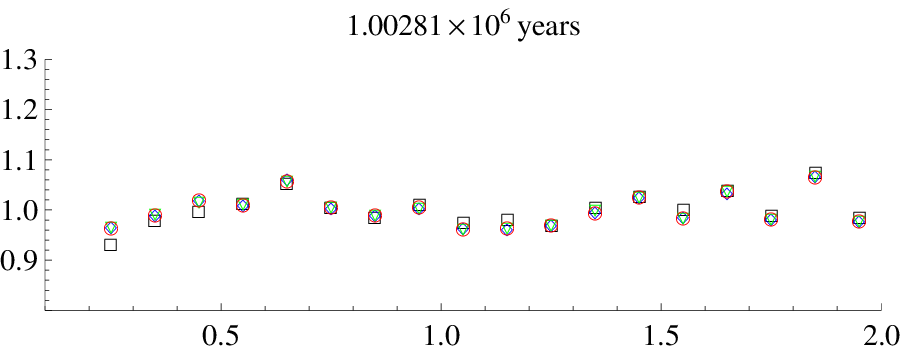}}\\

 \resizebox{0.48\hsize}{!}{\includegraphics[angle=0]{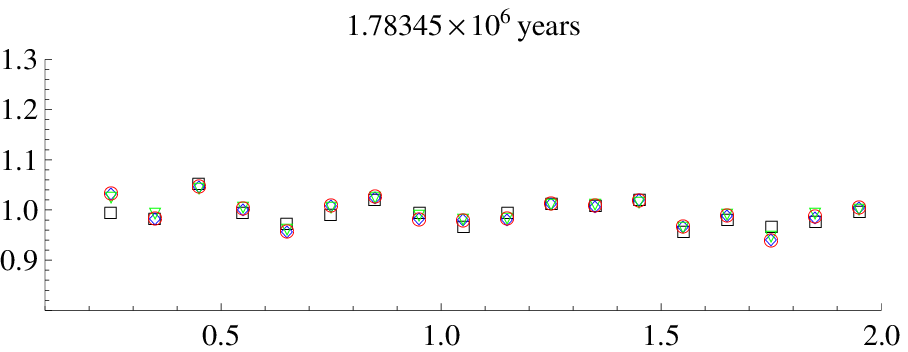}}
 \resizebox{0.48\hsize}{!}{\includegraphics[angle=0]{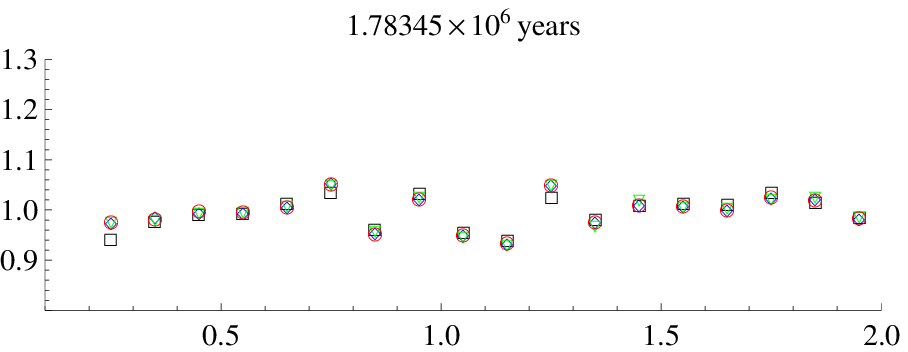}}\\

 \resizebox{0.48\hsize}{!}{\includegraphics[angle=0]{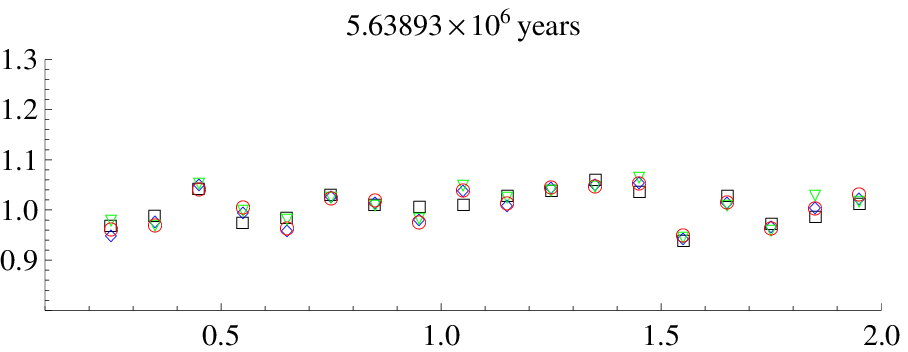}}
 \resizebox{0.48\hsize}{!}{\includegraphics[angle=0]{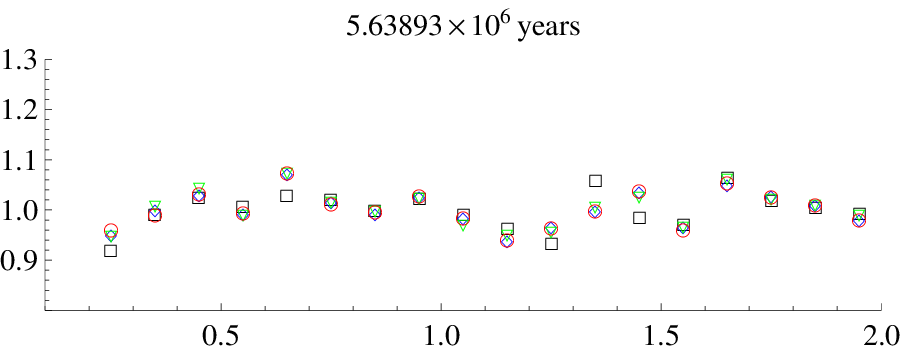}}\\

 \resizebox{0.48\hsize}{!}{\includegraphics[angle=0]{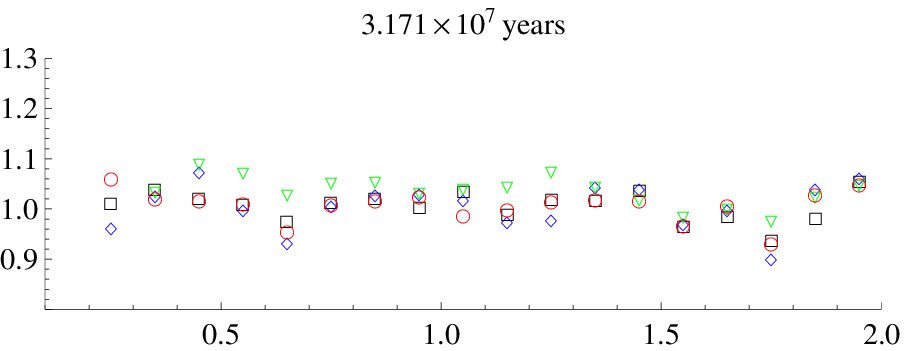}}
 \resizebox{0.48\hsize}{!}{\includegraphics[angle=0]{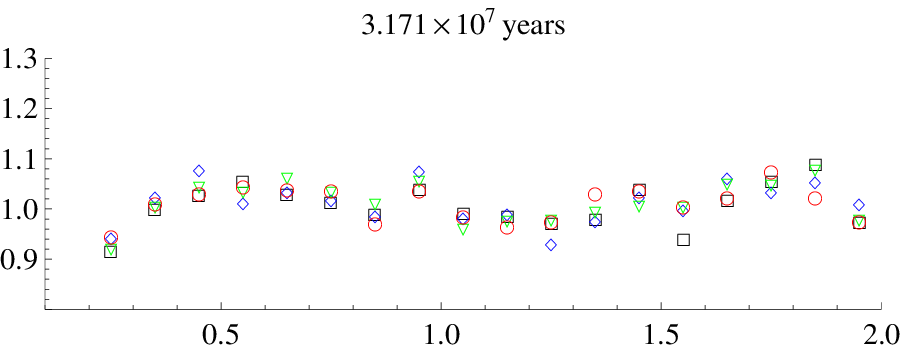}}\\

 \end{tabular}
\caption{The density profiles at $t=32.1\cdot 10^3 \text{years}=0.321$~{kyr}, $564$~{kyr},
$1.0\cdot 10^6 \text{years}=1$~{Myr}, $1.78$~{Myr}, $5.64$~{Myr}, and $31.7$~{Myr} (see the top
caption of each plot) for the main simulation with $10^5$ particles (left plots), and the auxiliary
simulation with the same number of particles ($10^5$), doubled softening length $\epsilon =
3.4$~{pc}, and the independently generated initial conditions (right plots). Black squares, red
circles, blue diamonds, and green triangles correspond to the direct summation (\texttt{ph4} code),
$\fac=5\cdot 10^{-5}$, $5\cdot 10^{-4}$, and $5\cdot 10^{-3}$, respectively.}
 \label{21fig4}
\end{figure*}

The second interesting effect appearing in the simulations is an early occurrence of an
instability. Figure~(\ref{21fig4}) represents the density profiles of the system at $t=32.1\cdot
10^3\text{years}=32.1$~{kyr}, $564$~{kyr}, $1.0\cdot 10^6\text{years}=1.0$~{Myr}, and $5.64$~{Myr}.
Distinct 'waves' in the density profiles occur already at $32.1$~{kyr} and reach the maximum
amplitude at $\sim 1.0$~{Myr}. After that, the amplitude of the instability does not change much:
the perturbations do not grow and do not disappear. The 'waves' are in evidence in
figure~(\ref{21fig1}) as well. It is remarkable that the instability does not differ for all four
values of $\fac$ that we used: the positions and the amplitudes of the 'waves' are identical. The
amplitude also does not depend on the softening parameter.

We may also make an another important conclusion from figure~(\ref{21fig4}): the instability has
nothing to do with the Poisson noise, or, more generally, any transient effects before the initial
conditions settle into an equilibrium configuration. First, the waves are just too strong, their
amplitude is $\sim 10\%$, while we may expect from equation~\ref{21e3} the amplitude $< 2\%$.
Second, the Hernquist profile is stable, and tidal effects and dynamical friction tend to destroy
any small substructures like the density waves \citep{bt}. As we could see in
section~\ref{poisson}, the overdensities with small density contrast are easily destroyed by tidal
effects. Nevertheless, the waves survive and even grow. Third, we performed the auxiliary
simulation with the same number of particles, but with the doubled softening length and
independently generated initial conditions (right panel in figure~(\ref{21fig4})). The positions of
the Poisson clumps are quite different in this case, but the waves look quite similar, and even the
positions of some of them are the same as in the main simulation. It suggests that the instability
does not depend either from the softening radius, nor from the Poisson noise in the initial
conditions.

We consider the behavior of the mean ($\mu$) and mean-square-root ($\sigma$) variations of the
integrals of motion, as we described in the previous section. It turns out that for all the
integrals at all radii the mean variation is, at least, one or (more typically) two orders of
magnitude smaller than the mean-square-root variation\footnote{Except of the trivial case, when the
core forms, and a lot of particles collectively leave the center. However, the core formation
occurs relatively late, and we can easily avoid the influence of this effect, just disregarding the
data after its beginning.}. It means that the simulation codes conserve the net angular momenta and
energy much better, than the characteristics of individual particles.

The temporal behavior of variations of $K$, $K_x$, $K_y$, $K_z$ turns out to be almost
indistinguishable (and, of course, $\sigma(\Delta K_x)\simeq\sigma(\Delta K_y)\simeq\sigma(\Delta
K_z)\simeq\sigma(\Delta K)/\sqrt3$). Therefore, we present the results only for $K$ instead of that
for $K_x$, $K_y$, $K_z$. What is more, the temporal behavior of all the five integrals is also very
similar for the \texttt{GADGET-2} and \texttt{ph4} simulations, especially the results for the tree
simulations with the \texttt{GADGET-2}, which are almost identical. Therefore, we plot only the
results for the direct summation (black crosses) and \texttt{GADGET-2} with $\fac=5\cdot 10^{-3}$,
since they differ the most.

\begin{figure*}
 \resizebox{\hsize}{!}{\includegraphics[angle=0]{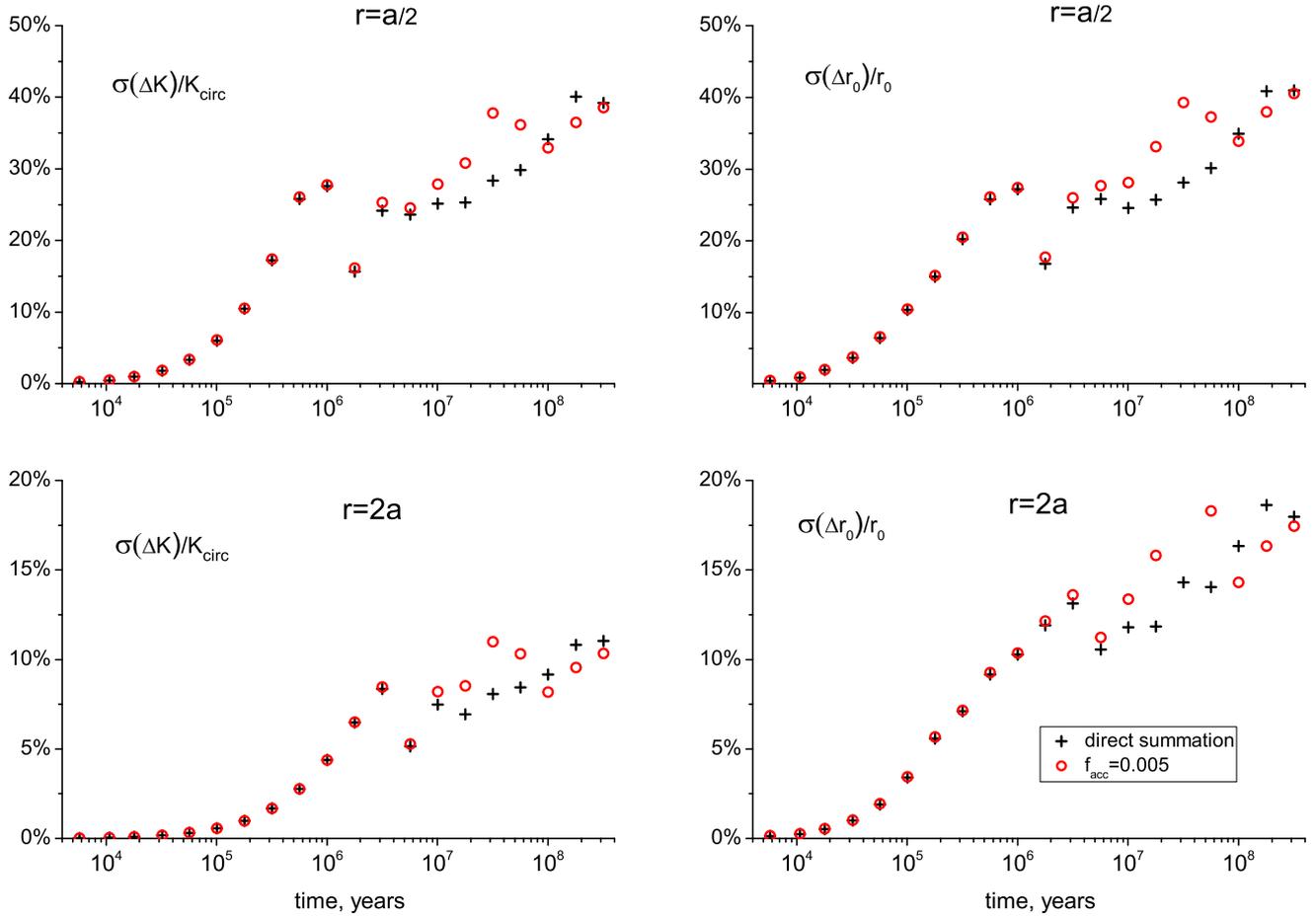}}
\caption{The temporal behavior of $\sigma(\Delta K)/K_{circ}$ (left panel) and $\sigma(\Delta
r_0)/\overline{r_0}$ (right panel) at the radii $r=a/2$ (upper panel) and $r=2a$ (lower panel) for
the direct summation algorithm (\texttt{ph4}, black crosses) and \texttt{GADGET-2} with
$\fac=5\cdot 10^{-3}$ (red circles). }
 \label{21fig5}
\end{figure*}

Figure~(\ref{21fig5}) represents the behavior of $\sigma(\Delta K)$ (left panel) and $\sigma(\Delta
r_0)$ (right panel) at the radii $r=a/2$ (upper panel) and $r=2a$ (lower panel). As we can see,
even the behaviors of $\sigma(\Delta K)$ and $\sigma(\Delta r_0)$ are quite similar: they linearly
grow with time, then reach their maximum at $t\sim \tau_d(r)$, then slightly decrease, and then
grow again, but significantly slower. The temporal behavior resembles that of the above-discussed
instability of the density profile (figure~(\ref{21fig4})), and it is more than proper to assume
that this is actually the same phenomenon.

Finally, the instability is quite distinct in figures~(\ref{21fig8}) and~(\ref{21fig11})
representing the upward $N_{+}(t)$ (black crosses) and downward $N_{-}(t)$ (red circles) radial
streams of particles through the spheres\footnote{See section~\ref{21sec3} for exact definitions of
the streams.} of radii $r=a/2$ and $r=2a$, respectively, divided by the total number of the
particles $N(r)$ inside $r$. The separation of the cross and circle lines at the late parts of the
plots corresponds to the core formation and has nothing to do with the instability. However, other
features of the plot closely resemble these of the figure~(\ref{21fig5}): $N_{+}(t)$ and $N_{-}(t)$
(the number of particles, which apocenter distances ${r_0}$ have crossed the spheres of radii
$r=a/2$ and $r=2a$) rapidly grow with time, then reach their maximum at $t\sim \tau_d(r)$, then
slightly decrease, then some particles even return back ($N_{+}(t)$ and $N_{-}(t)$ decrease), and
then grow again, but significantly slower.

\clearpage

The behavior of the quantities presented in figures~(\ref{21fig5}), (\ref{21fig8}),
and~(\ref{21fig11}) clearly indicates that the rapid initial variations of the integrals of motion
are not due to a slow accumulation of numerical inaccuracies or the collisional relaxation: first,
the decrease of the standard deviations (figure~\ref{21fig5}), as well as the decrease of the
number of particles having crossed the radii $r=a/2$ (figure~\ref{21fig8}) and $r=2a$
(figure~\ref{21fig11}), at $t\sim \tau_d(r)$ would be impossible in this case. Second, the standard
deviations in figure~(\ref{21fig5}) grows linearly for $t< \tau_d(r)$, which suggests that we deal
with a collective instability, and not with any sort of a stochastic process. Indeed, in
figure~\ref{21fig4} we can see the appearance of density waves with the amplitude corresponding to
that of the integrals variations.

We should emphasize that \emph{the instability is undeniably a numerical effect}. The
Doremus-Feix-Baumann theorem together with the Antonov's second law (see \citep{bt} for details)
prove: the Hernquist profile with the isotropic velocity distribution that we model is stable.

What could be the influence of the instability on the cosmological simulations, in particular, the
simulations of the hierarchical structure formation? We will discuss this question in more details
in the next subsection, here restricting ourselves to a remark that the effect that we have found
resembles the violent relaxation \citep{violent} (hereafter we abbreviate it as VR), which may
occur during the halo formation from an initial small perturbation. The essence of later process is
simple: when the halo collapses, strong and rapidly evolving density inhomogeneities (caustics
etc.) should appear. The inhomogeneities create a small-scale gravitational field, and as a result
the halo particles may effectively exchange their energies and other integrals of motion. The
characteristic time of the violent relaxation $t\sim \tau_d(r)$ coincides with that of the
instability that we have found: the violent relaxation 'works' only during the halo collapse.

However, the real violent relaxation has nothing to do with the numerical 'violent relaxation' that
we have found. The violent relaxation is a quite real effect that may occur (but not necessarily
occurs) during the halo formation. It certainly cannot take place in our simulations, since we
consider a stationary and stable profile. There are also less fundamental differences between the
effects: for instance, the efficiency of the violent relaxation rapidly drops with radius, while
figure~(\ref{21fig4}) shows that the numerical 'violent relaxation' is rather effective even at
large radii.

A major challenge for N-body simulations is that there seems to be no apparent way to separate the
real and numerical violent relaxations in realistic cosmological modelling. Furthermore, as we will
see, the numerical relaxation in simulations of the hierarchical structure formations may, in all
likelihood, be far in excess of the $\sim 10\%$ that we obtain for the stationary halo and
completely transfigure the energy portrait of the system.

We may say a few words about the nature of the instability. Since its development is not sensitive
to $\fac$, the instability does not depend on the algorithm of gravitational field evaluation. It
is hardly an effect of discretization: if this were so, the effect would drastically depend of
radius (as, for instance, the collisional relaxation does). The phenomenon is probably produced
either by an inadequate accuracy of the trajectory evaluation algorithm, or by the Miller's
instability.

Actually, it is the Miller's instability (together with the calculation time) that makes an
employment of precise and heavy algorithms of trajectory evaluation in N-body simulations low
useful: the Miller's instability makes the Liapunov time comparable with the dynamical time of the
system \citep{miller1964}. Even if we take into account the specificity of N-body algorithms (like
the potential smoothing), the instability arises in a time, which is much shorter than $\tau_r(r)$
at the given radius and remains comparable with the dynamical time $\tau_d(r)$
\citep{valluri2000,hut2002}. The credence to the results of N-body simulations under these
conditions is totally based on the resulting density profile stability. Indeed, different N-body
codes, with various algorithms of potential and trajectory evaluation lead to quite similar
profiles of the formed halos. Relying on this, the profile is considered to be physically
meaningful and describing real halos, despite of the fact that the orbits of individual particles
have no physical significance \citep[section 4.7.1(b)]{bt}. Now we try to show on the basis of our
simulations, how vulnerable this reasoning is.

\subsection{The variations of the integrals of motion and the 'core-cusp problem'}

\subsubsection{Are the variations of integrals of motion of individual particles important?}

Thus we have found that, on the one hand, the integrals of motion of individual particles
significantly vary in a short time (they change on $\sim 10\%$ in $\sim 10^6$~{years}, i.e., $\sim
10^{-4}$ of the age of the Universe). On the other hand, the density profile is, generally
speaking, quite stable and does not decline much from the analytical solution. This raises a very
important question. It is widely believed that, though the Millers's instability totally ruins the
orbits and the integrals of motion of individual particles, the statistical properties of the
particle distribution (in particular, the density profiles) are trustable. Indeed, N-body
simulations of the clustering of dark matter lead to a universal density profile of the halos
(hereafter we abbreviate it as UDP), low-sensitive to the simulation parameters \citep{power2003}.
Unfortunately there is no adequate analytical model of the clustering, and we cannot compare the
simulation results with the analytical solution. However, the fact that different computer codes
lead to very similar UDPs (though the UDP itself gradually changes from the NFW in 1997 to the
Einasto profile with $n\simeq 6$ in 2017) creates an impression that the DM density profiles may be
reliably simulated despite of inaccuracies of the velocity distribution and any dynamical
parameters of individual particles \citep{gao2008, dutton2014}.

Of course, this is not true. First, the velocity distribution is mutually bound with the density
profile. For simplicity, let us consider a stationary spherically symmetric DM halo with an
anisotropic velocity distribution of the particles. Then the particle distribution $f$ in the phase
space is a function of only the particle energy $w$ \citep{bt}: $f(r,\vec v)\equiv f(w)=
f(\phi(r)+v^2/2)$. The velocity distribution and the density at some radius $r$ are equal to
$f(\phi(r)+v^2/2) dv^3$ and $\int 4\pi v^2 f(\phi(r)+v^2/2) dv$, respectively. Thus, any inaccuracy
of the velocity distribution directly translates into an inaccuracy of the density profile.

The second, more fundamental argument is the following. We consider a spherically symmetric halo;
therefore, we may use the integrals of motion $K$, $K_x$, $K_y$, $K_z$, and $r_0$ as a full set of
generalized coordinates\footnote{Actually, we may even get rid of one of the integrals $K$, $K_x$,
$K_y$, $K_z$, since they are bound by the trivial equation $K^2=K^2_x+K^2_y+K^2_z$.} and consider
the distribution function $f$ of the system as a function of only these five coordinates. A
collisionless system (for instance, a DM system, which is supposed to be collisionless) obeys the
collisionless kinetic equation $df/dt=0$. On the other hand, if the integrals of motion experience
a relatively slow evolution (which is the case), the system may be described by the Fokker-Planck
equation \citep{ll10}
\begin{equation}
\dfrac{df}{dt}=\frac{\partial}{\partial q_\alpha}\left\{{A_\alpha} f+\frac{\partial}{\partial
q_\beta}[B_{\alpha\beta}f]\right\}
 \label{fokker_planck}
\end{equation}
where $q_\alpha$ are the integrals of motion, and $A$ and $B$ are the Fokker-Planck coefficients.
\begin{equation}
{A_\alpha}=\dfrac{\overline{\delta q_\alpha}}{\delta t}\qquad
B_{\alpha\beta}=\dfrac{\overline{\delta q_\alpha\delta q_\beta}}{2\delta t}.
 \label{fokker_planck_coefficients}
\end{equation}
Figure~(\ref{21fig5}) illustrates that at least some of $B$s tangibly differ from zero. Hence the
N-body system in all our simulations behave as essentially collisional.

Thus, a reliable simulation of DM density profiles without a reliable simulation of the velocity
distribution and the integrals of motion of individual particles is out of the question.

\begin{figure*}
 \resizebox{\hsize}{!}{\includegraphics[angle=0]{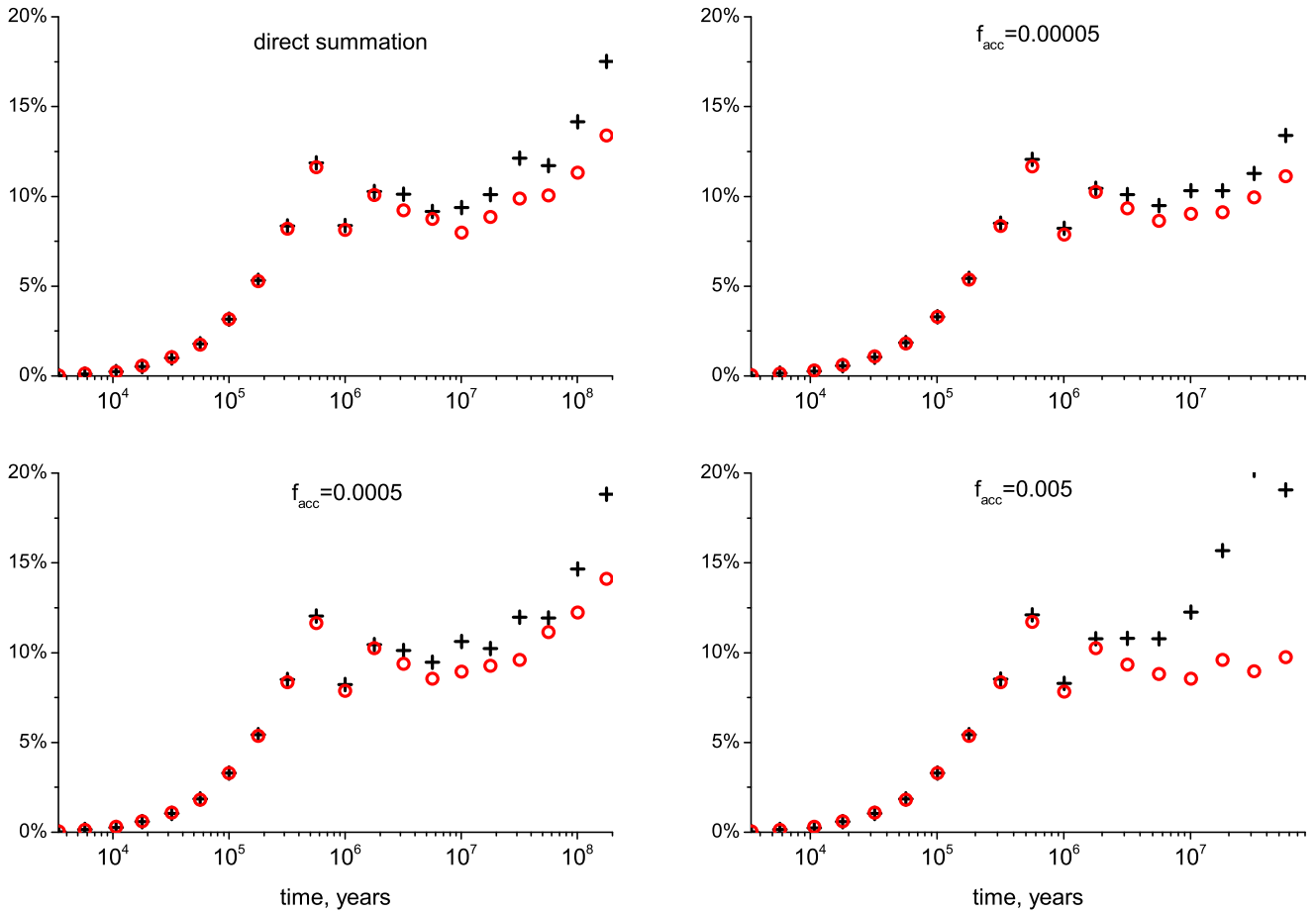}}
\caption{The upward $N_{+}(t)$ (black crosses) and downward $N_{-}(t)$ (red circles) radial streams
of particles (see section~\ref{21sec3} for exact definition) through the sphere of radius $r=a/2$,
divided by the total number of the particles $N(a/2)$ inside $a/2$.}
 \label{21fig8}
\end{figure*}

\subsubsection{The 'core-cusp problem' and the halo relaxation}

The 'core-cusp problem' gives us an apt illustration of the importance of correct modelling of the
behavior of the integrals of motion (in particular, of the particle energies) in simulations for
reliability of their results. Indeed, DM halos appear from small linear cosmological perturbations.
It is easy to show analytically  that a strong relaxation is absolutely necessary during the halo
formation in order to form a cuspy profile; the resulting profile is obligatory cored if the
relaxation is moderate or weak \citep{15,16}. Fundamentally, the only plausible mechanism of the
strong relaxation during the formation of a DM halo is the violent relaxation\footnote{In
principle, barionic matter may also participate in the relaxation of real astrophysical halos.
However, we discuss the 'core-cusp problem', which is the most acute just in the case of halos that
contain an unusually small fraction of barionic matter, like dwarf spheroidals \citep{walker2011}.
Therefore, it is unlikely that taking account of the barionic matter influence may resolve the
'core-cusp problem' \citep{diversity}.}. However, it is important to understand, how effective the
real VR is, i.e., how significantly the initial specific energy is redistributed between the DM
particles during the formation of an astrophysical halo. The task of the gravitational collapse is
too complex to obtain an exact analytical solution for realistic situations. Observations rather
suggest that the relaxation is not violent \citep{16}. And our simulations show that a numerical
'violent relaxation' occurs in the N-body modelling, which may be easily confused with the real VR.

There is no apparent way to control the integrals of motion in real cosmological simulations, but
we model the Hernquist profile, which is very close the universal density profile obtained in
cosmological simulations. We find the change of the integrals on $\sim 10\%$ in the time interval
of $\sim 10^{-4}$ of the age of the Universe. Though the integrals of motion evolve much slower
after the initial 'violent relaxation', we should emphasize that the numerical artefacts in
simulations of the hierarchical structure formations may, in all likelihood, be far in excess.
First, we model an ideal system, a stationary halo. Second, the effects of the 'numerical
interaction' between particles occurring in complex simulations of the hierarchical clustering are
most likely to be more perceptible: the simulation inaccuracies may be larger during the
highly-nonstationary process of halo collapse. Third, the numerical 'violent relaxation' may occur
over and over again in repeating merging of smaller halos into a larger one during the hierarchical
clustering. Fourth, the first halos always contain only few tens of particles, and later these
first halos contribute to the cusps of more massive ones. We confront with a question: are the
halos with small $N$ simulated properly? This problem has been discussed in literature in the
context of two-body relaxation \citep{binney2002}, but if the numerical effects are already
important at the dynamical time, the problem of the first halos may be much more important.

Hence the N-body simulations also give no plausible way to determine the efficiency of the
relaxation occurring during the formation of astrophysical DM halos. A strong relaxation is
absolutely necessary to form a cuspy profile from the initial small perturbation \citep{15,16}, and
it apparently occurs in N-body simulations. However, if the relaxation is real, the cusps routinely
occurring in the centers of DM halos in cosmological N-body simulations correspond to the
properties of real astrophysical systems. If the relaxation is caused by numerical effects --- the
cusps are no more than a numerical artefact. In the first case, the 'core-cusp problem' gives us
the extremely valuable information that cold DM paradigm is wrong: the DM should be either warm, or
self-interacting, or has some other non-trivial physical properties. In the second case, we still
need to improve N-body codes in order to be able to make physical conclusions from discrepancies of
the simulation results and observations.

It is a wide-spread opinion (based mainly on \citep{power2003} and similar works) that the cusp
formation in the centers of DM halos is a genuine property of collision-less systems: indeed, in
N-body simulations it forms in a time, which is comparable with $\tau_d(r)$ and much shorter than
$\tau_r(r)$. Meanwhile, the core formation at $t\simeq 1.7 \tau_r(r)$ is considered as the first
sign of relaxation. Thus, it is believed that the relaxation tends to destroy the cusp.

Our simulations show that both these statements are doubtful. First, the core formation at $t\simeq
1.7 \tau_r$ is caused by inaccuracy of the tree algorithm, and not by a relaxation. Second, the
density profiles corresponding to the direct summation code \texttt{ph4} are almost identical in
the upper and lower panels of figure~(\ref{21fig1}), i.e., at time moments $t=0.45$~{Gyr} and
$t=2.85$~{Gyr}. There is no core formation, and the profile for $r\ge 0.2a$ has little difference
with the initial one. The small depression at $r\le 0.1a$ is most likely not the core, but an
effect of the potential softening: its radius is $\sim 6\epsilon$, and (contrary to the core) it
does not grow with time. Meanwhile, the upper and lower panels of figure~(\ref{21fig1}) correspond
to $\sim 1.4$ and $9 \tau_r$ at $r= 0.5 a$ and $\sim 11$ and $70 \tau_r$ at $r= 0.2 a$,
respectively. The influence of, at least, the collisional relaxation should be very significant (or
even decisive) at $t\sim 70 \tau_r$. However, we can see no cusp destruction outside $r= 0.1 a$
(contrary to the \texttt{GADGET-2} simulations). Thus, the relaxation effects do not obligatory
tend to transform a cusp into a core. On the other hand, as we have already mentioned, a strong
relaxation is necessary to form a cuspy profile from the initial small perturbation. Hence
relaxation processes tend to form the cusp, rather than to destroy it.

\begin{figure*}
 \resizebox{\hsize}{!}{\includegraphics[angle=0]{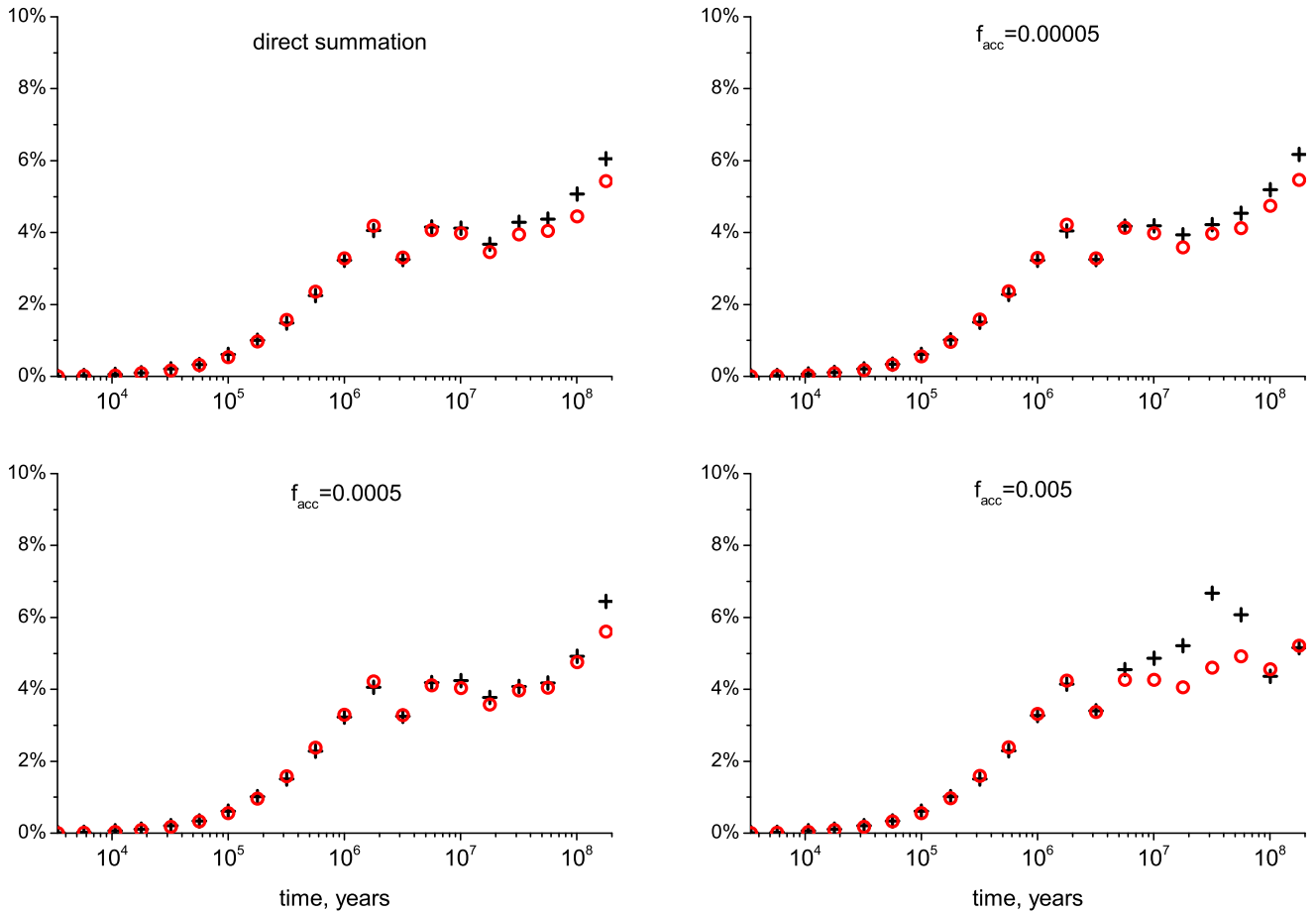}}
\caption{The upward $N_{+}(t)$ (black crosses) and downward $N_{-}(t)$ (red circles) radial streams
of particles (see section~\ref{21sec3} for exact definition) through the sphere of radius $r=2a$,
divided by the total number of the particles $N(2a)$ inside $2a$.}
 \label{21fig11}
\end{figure*}

\subsubsection{The Fokker-Planck streams}

The standpoint that the cusps in the halo centers is a numerical artefact is strongly supported by
the properties of the upward $N_{+}(t,r)$ and downward $N_{-}(t,r)$ radial drifts of particles in
the halo. As one can see in figure~(\ref{21fig8}), $\sim 12\%$ of particles inside $r=a/2$ cross
this radius in a quite short period of time $t\sim \tau_d$ (to be more precise, their apocenter
distances $r_0$ cross $r=a/2$), and the halo profile is stable only because the upward and downward
streams compensate each other. Indeed, the curves in figures~(\ref{21fig8}) and (\ref{21fig11})
branch off at $\sim 10$~{Myr}, earlier in figure~(\ref{21fig8}) and later in
figure~(\ref{21fig11}). The appearing imbalance between the upward and downward streams corresponds
to the core formation. The divergence of the curves for the \texttt{ph4} simulation (the upper left
panels in both the figures) is rather small, which depicts the strong suppression of the core
formation in this instance. It is interesting to mention that the $N_{+}$ and $N_{-}$ curves branch
off much earlier, than the moment when the core becomes visible in the density profile.

Let us underline several important points. First, the drifts $N_{+}$ and $N_{-}$ are purely
numerical effects: $N_{+}(t,r)=N_{-}(t,r)\equiv 0$ in the collisionless case, since $r_0$ is an
integral of motion. Second, the streams are not just a minor effect of a low-amplitude particle
wiggling: as we have already mentioned, each particle is counted no more than once in calculation
of $N_{+}$ and $N_{-}$. What is more, the fluxes are just too strong. For instance, the number of
particles crossing the radius $r=a/2$ in $t\simeq\tau_d(a/2)$ (figure~(\ref{21fig8})) is
approximately equal to the total number of particles between $a/2$ and $a/2 + 0.04a$. Thus, the
orbits of particles move as a whole in the halo due to some significant numerical effects, and the
amplitude of this drift is quite large on average.

It may be argued that, though the numerical streams under consideration are rather intensive, they
are insufficient to determine the profile shape, since only $\sim 12\%$ of particles inside $r=a/2$
are carried through this sphere in $t\sim \tau_d(a/2)$, and then the streams significantly weaken.
This is not so. First, the upper left panel in figure~(\ref{21fig8}) (the only panel in the figure,
where the influence of the core formation is relatively small) shows that $N_{+}$ and $N_{-}$ are
still growing after $t\sim \tau_d$, even if much slower than during the numerical 'violent
relaxation'. They reach $\sim 15-20\%$ at $t=0.2$~{Gyr}, and if their growth continues like this,
they may reach $\sim 100\%$ at $t=1.36$~{Gyr}. It means that the numerical effects may thoroughly
redistribute the particles even inside the radius $r=a/2=r_s$. Moreover, this radius is very large,
the 'core-cusp problem' occurs much closer to the center, while the efficiency of the numerical
streams grows in inverse proportion of radius.

We cannot reliably calculate $N_{+}(t,r)$ and $N_{-}(t,r)$ for $r\ll a/2$, because the core
formation starts too early at small radii. Since we use the slow algorithm \texttt{ph4}, we have to
limit ourselves with $10^5$ particles in the halo. It spoils the spacial resolution, and the
relaxation time becomes comparable with the dynamical one as we approach the halo center. However,
we can estimate the behavior of $N_{+}(t,r)$ and $N_{-}(t,r)$ close to the halo center if we
combine our results with those of \citep{17}. These authors also simulate the Hernquist halo, but
they test only the \texttt{Gadget-2} code with $\fac=0.005$. As a result, simulations \citep{17}
contain much more particles and gain significantly higher spatial resolution. The authors of
\citep{17} also found the unphysical streams $N_{+}(t,r)$ and $N_{-}(t,r)$, and in the area where
we can compare the results, they coincide with ours. In the region between $r=0.1 a$ and $r=0.5 a$
the ratios $N_{+}(\tau_d,r)/N(r)$ and $N_{-}(\tau_d,r)/N(r)$ (i.e., the number of particles crossed
the sphere of radius $r$ upward and downward, respectively, in the dynamical time at this radius,
divided by the total number of the particles $N(r)$ inside $r$) are in inverse proportion of $r$
(see figure~3 in \citep{17}). It means that the fraction of particles carried by the unphysical
streams through the sphere of radius $r=0.1 a=0.2 r_s$ in $t\sim \tau_d$ reaches $\sim 60\%$ of the
total mass inside the sphere. Recall that we consider the ideal instance of a stationary and stable
halo. If in the cosmological simulations of hierarchical structure formation the balance between
the upward and downward streams is violated, the $60\%$ mass drift is sufficient to transfigurate
the profile entirely, in particular, to form the cusp in $t\sim \tau_d$.

Hence a purely numerical effect, the upward and downward streams of particles, are strong enough to
totally determine the shape of the density profile in the centers of cosmological halos. The
profile is stable not due to the absence of numerical effects, but just due to the mutual
compensation of two strong numerical artefacts. Moreover, we do not know for sure, why the upward
$N_+$ and downward $N_-$ streams initially compensate each other so well, and why the
self-compensation ends at some moment leading to the core formation. One can see that the last
process is defined by the parameters of N-body codes, having nothing to do with the properties of
real DM systems. Besides this, the time of the numerical 'violent relaxation' that we found ($t\sim
\tau_d$) coincides with the characteristic time ($\sim \tau_d$) of the central cusp formation in
cosmological N-body simulations.

\subsubsection{Resume}

To summarize, the unphysical stream of particles $N_-$ during the numerical 'violent relaxation'
can bring through the radius $r=0.1a=0.2 r_s$ approximately $60\% $ of the mass that the resulting
halo contains inside this radius. This is more than enough to form the cusp in the halo center in
$t\sim \tau_d$ by purely numerical effects. The intensive variations of the integrals of motion of
individual particles reveals a sort of interaction between them (it does not matter if it is
physical or numerical), and the system is described not by the collision-less kinetic equation, but
by the Fokker-Planck one. The Fokker-Planck equation has its own stationary solutions, and the
universal density profile obtained in simulations falls exactly on the stationary solution of the
Fokker-Planck equation in the halo center \citep{18,13}. All these facts suggest that the cusp
routinely occurring in the centers of DM halos in cosmological simulations is no more than a
numerical artefact.

Finally, the 'core-cusp problem' is the most known, but not the only strange phenomenon occurring
in N-body simulations. The authors of \citep{vanderbosch2018, vanderbosch2018b} report that, in the
absence of baryonic processes, the complete physical disruption of CDM substructure should be
extremely rare\footnote{Analytical consideration supports this statement \citep{19}.} and that most
disruption in numerical simulations are, most likely, artificial. The authors suppose that
subhaloes in N-body simulations suffer from an instability triggered by the amplification of
discreteness noise in the presence of a tidal field. We can see, however, that some numerical
instability occurs even in a stationary and stable halo, and even in the absence of any tidal
field.

The main conclusion that one can make from these facts is that the present state-of-art of N-body
simulation tests is absolutely insufficient. First, the criteria of reliability based on the
profile stability and universality are extremely questionable and may lead to false convergence, as
we could see. The full set of integrals of motion of the system should be considered to reveal
undesirable numerical effects with assurance. Yes, precise evaluation of trajectories of each
particle is probably unnecessary to reliably reproduce the halo density profiles. However, the
neglect of the energy and momentum exchange between particles is surely not harmless.

Second, one can see that the density profile in \texttt{GADGET-2} simulations is unresponsive to
decreasing of $\fac$ below $5\cdot 10^{-4}$. Thus, we reach a sort of 'convergence', but the
convergence is false: the behavior of the real DM system that we model (or even of the system where
the gravitational force is evaluated with the help of the direct summation algorithm) is
significantly different. This is a good illustration why a comparison of the simulation results
with the analytical model should always be preferred over a comparison of the simulation results
with another simulation results as a reliability test. Indeed, an analytical solution of the
extremely complex task of the hierarchical structure formation is scarcely possible. However, it is
usually possible to offer an analytical model close enough to the realistic one. For instance, for
each halo obtained in simulations, an analytical model of an isolated halo with similar profile and
similar velocity distribution may always be obtained \citep{bt}, which allows to test the
simulation properties comprehensively, with the use of the full set of dynamical parameters of the
system. Only irreproachably reliable estimations of the simulation accuracy and convergence may
permit to make strong physical conclusions: otherwise any discrepancy with observations may turn
out to be just a result of an unaccounted numerical effect.

\section{Summary}
\label{21sec5}
\begin{enumerate}
 \item
The reliability criteria of the N-body simulations based on the profile stability and universality
(like \citep{power2003}) are unsafe. In particular, the opinion that the universal density profile
(UDP) commonly occurring in the cosmological simulations corresponds to the absence of any
significant numerical effects, while the core formation in the center is a result and the first
sign of the collisional relaxation, is absolutely incorrect. First, very essential numerical
effects occur much earlier than the core formation. Second, the core formation has nothing to do
with the collisional relaxation, being defined by the parameters of the tree algorithm. Thus, the
criteria like $t\le 1.7 \tau_r(r)$ are based on artificial binding of two essentially independent
processes and therefore cannot be valid.

 \item
An instability with the characteristic time $\sim \tau_d(r)$ develops immediately after the
simulation launch. It leads to a numerical 'violent relaxation': the integrals of motion change (on
the average) on $10\%$ from their initial values even at $r\simeq r_s$. In contrast to the real
violent relaxation \citep{violent}, the relaxation that we found is purely numerical. The mechanism
of its occurrence is not quite clear; most likely, this is a result of either an inadequate
accuracy of the trajectory evaluation algorithm, or the Miller's instability.
 \item
Relying on the present-day N-body simulations, one cannot infer that a relaxation (in particular,
the collisional one) tends to transform a cusp into a core in the center of DM halos. Theoretical
consideration rather suggests the opposite: a relaxation assists to the cusp formation, at least,
during the halo formation.
 \item
The necessity of a strong relaxation for a cuspy profile formation \citep{15, 16} makes critical
the issue of correct modelling of the initial halo relaxation in cosmological simulation. If the
N-body simulations overestimate the relaxation during the halo formation (as a result of the
numerical 'violent relaxation' that we found), it may lead to a false cusp formation in the halo
center. Our results give every reason to believe that it is exactly what happens in the N-body
simulations of the large-scale structure formation. Then the 'core-cusp problem' is no more than a
technical problem of N-body simulations.

 \item
The significant variations of the integrals of motion reveal that the system of test particles in
the N-body simulations is essentially collisional, contrary to real DM systems. In the idealized
case of a stationary and stable halo that we consider, the variations do not affect much the
density profile, but their influence cannot be minor in simulations of the hierarchical structure
formation of the Universe.

 \item
Much remains to be done in testing of N-body simulation convergence and reliability. A comparison
of the simulation results with the analytical model should always be preferred over a comparison of
results of different simulations.

\end{enumerate}

\section{Acknowledgements}

The work is supported by the Program 28 of the fundamental research of the Presidium of the Russian
Academy of Sciences ``Space: research of the fundamental processes and relations'', subprogram II
``Astrophysical objects as space laboratories''. The study is supported by the Supercomputing
Center of Lomonosov Moscow State University. The study is supported by the "GOVORUN" Supercomputer,
the Bogoliubov Laboratory of Theoretical Physics and Laboratory of Information Technologies, Joint
Institute for Nuclear Research, Dubna, Russia.

This research is supported by the Munich Institute for Astro- and Particle Physics (MIAPP) of the
DFG cluster of excellence "Origin and Structure of the Universe".

















\begin{thebibliography}{40}
\expandafter\ifx\csname natexlab\endcsname\relax\def\natexlab#1{#1}\fi

\bibitem[{{Athanassoula} {et~al.}(2000){Athanassoula}, {Fady}, {Lambert}, \&
  {Bosma}}]{2000MNRAS.314..475A}
{Athanassoula}, E., {Fady}, E., {Lambert}, J.~C., \& {Bosma}, A. 2000, \mnras,
  314, 475

\bibitem[{{Barnes} \& {Hut}(1986)}]{tree1}
{Barnes}, J. \& {Hut}, P. 1986, \nat, 324, 446

\bibitem[{{Baushev}(2014{\natexlab{a}})}]{15}
{Baushev}, A.~N. 2014{\natexlab{a}}, \apj, 786, 65

\bibitem[{{Baushev}(2014{\natexlab{b}})}]{16}
{Baushev}, A.~N. 2014{\natexlab{b}}, \aap, 569, A114

\bibitem[{{Baushev}(2015)}]{13}
{Baushev}, A.~N. 2015, Astroparticle Physics, 62, 47

\bibitem[{{Baushev}(2016)}]{19}
{Baushev}, A.~N. 2016, \jcap, 1, 018

\bibitem[{{Baushev} \& {Barkov}(2018)}]{18}
{Baushev}, A.~N. \& {Barkov}, M.~V. 2018, \jcap, 2018, 034

\bibitem[{{Baushev} {et~al.}(2017){Baushev}, {del Valle}, {Campusano},
  {Escala}, {Mu{\~n}oz}, \& {Palma}}]{17}
{Baushev}, A.~N., {del Valle}, L., {Campusano}, L.~E., {et~al.} 2017, \jcap, 5,
  042

\bibitem[{{Binney} \& {Knebe}(2002)}]{binney2002}
{Binney}, J. \& {Knebe}, A. 2002, \mnras, 333, 378

\bibitem[{{Binney} \& {Tremaine}(2008)}]{bt}
{Binney}, J. \& {Tremaine}, S. 2008, {Galactic Dynamics: Second Edition}
  (Princeton University Press)

\bibitem[{{Chemin} {et~al.}(2011){Chemin}, {de Blok}, \& {Mamon}}]{mamon2011}
{Chemin}, L., {de Blok}, W.~J.~G., \& {Mamon}, G.~A. 2011, \aj, 142, 109

\bibitem[{{de Blok}(2010)}]{corecuspreview}
{de Blok}, W.~J.~G. 2010, Advances in Astronomy, 2010, 789293

\bibitem[{{Dehnen}(2001)}]{dehnen01}
{Dehnen}, W. 2001, \mnras, 324, 273

\bibitem[{{Dutton} \& {Macci{\`o}}(2014)}]{dutton2014}
{Dutton}, A.~A. \& {Macci{\`o}}, A.~V. 2014, \mnras, 441, 3359

\bibitem[{{Evans} \& {Collett}(1997)}]{evans1997}
{Evans}, N.~W. \& {Collett}, J.~L. 1997, \apjl, 480, L103

\bibitem[{{Gao} {et~al.}(2008){Gao}, {Navarro}, {Cole}, {Frenk}, {White},
  {Springel}, {Jenkins}, \& {Neto}}]{gao2008}
{Gao}, L., {Navarro}, J.~F., {Cole}, S., {et~al.} 2008, \mnras, 387, 536

\bibitem[{{Harvey} {et~al.}(2017){Harvey}, {Courbin}, {Kneib}, \&
  {McCarthy}}]{clustercores}
{Harvey}, D., {Courbin}, F., {Kneib}, J.~P., \& {McCarthy}, I.~G. 2017, \mnras,
  472, 1972

\bibitem[{{Hayashi} {et~al.}(2003){Hayashi}, {Navarro}, {Taylor}, {Stadel}, \&
  {Quinn}}]{hayashi2003}
{Hayashi}, E., {Navarro}, J.~F., {Taylor}, J.~E., {Stadel}, J., \& {Quinn}, T.
  2003, \apj, 584, 541

\bibitem[{{H{\'e}non}(1971)}]{nbodyunits}
{H{\'e}non}, M.~H. 1971, \apss, 14, 151

\bibitem[{{Hernquist}(1990)}]{hernquist1990}
{Hernquist}, L. 1990, \apj, 356, 359

\bibitem[{{Hernquist} \& {Katz}(1989)}]{tree2}
{Hernquist}, L. \& {Katz}, N. 1989, \apjs, 70, 419

\bibitem[{{Hut} \& {Heggie}(2002)}]{hut2002}
{Hut}, P. \& {Heggie}, D.~C. 2002, Journal of Statistical Physics, 109, 1017

\bibitem[{{Klypin} {et~al.}(2013){Klypin}, {Prada}, {Yepes}, {Hess}, \&
  {Gottlober}}]{klypin2013}
{Klypin}, A., {Prada}, F., {Yepes}, G., {Hess}, S., \& {Gottlober}, S. 2013,
  ArXiv e-prints

\bibitem[{{Knebe} {et~al.}(2000){Knebe}, {Kravtsov}, {Gottl{\"o}ber}, \&
  {Klypin}}]{knebe2000}
{Knebe}, A., {Kravtsov}, A.~V., {Gottl{\"o}ber}, S., \& {Klypin}, A.~A. 2000,
  \mnras, 317, 630

\bibitem[{{Landau} \& {Lifshitz}(1980)}]{ll10}
{Landau}, L.~D. \& {Lifshitz}, E.~M. 1980, {Statistical physics. Pt.1, Pt.2}

\bibitem[{{{\L}okas}(2002)}]{draco2002}
{{\L}okas}, E.~L. 2002, \mnras, 333, 697

\bibitem[{{Lynden-Bell}(1967)}]{violent}
{Lynden-Bell}, D. 1967, \mnras, 136, 101

\bibitem[{{Merritt}(1996)}]{merritt96}
{Merritt}, D. 1996, \aj, 111, 2462

\bibitem[{{Miller}(1964)}]{miller1964}
{Miller}, R.~H. 1964, \apj, 140, 250

\bibitem[{{Oman} {et~al.}(2015){Oman}, {Navarro}, {Fattahi}, {Frenk}, {Sawala},
  {White}, {Bower}, {Crain}, {Furlong}, {Schaller}, {Schaye}, \&
  {Theuns}}]{diversity}
{Oman}, K.~A., {Navarro}, J.~F., {Fattahi}, A., {et~al.} 2015, \mnras, 452,
  3650

\bibitem[{{Portegies Zwart} {et~al.}(2009){Portegies Zwart}, {McMillan},
  {Harfst}, {Groen}, {Fujii}, {Nuall{\'a}in}, {Glebbeek}, {Heggie}, {Lombardi},
  {Hut}, {Angelou}, {Banerjee}, {Belkus}, {Fragos}, {Fregeau}, {Gaburov},
  {Izzard}, {Juri{\'c}}, {Justham}, {Sottoriva}, {Teuben}, {van Bever},
  {Yaron}, \& {Zemp}}]{amuse1}
{Portegies Zwart}, S., {McMillan}, S., {Harfst}, S., {et~al.} 2009, \na, 14,
  369

\bibitem[{{Power} {et~al.}(2003){Power}, {Navarro}, {Jenkins}, {Frenk},
  {White}, {Springel}, {Stadel}, \& {Quinn}}]{power2003}
{Power}, C., {Navarro}, J.~F., {Jenkins}, A., {et~al.} 2003, \mnras, 338, 14

\bibitem[{{Springel}(2005)}]{springel2005}
{Springel}, V. 2005, \mnras, 364, 1105

\bibitem[{{Springel} {et~al.}(2001){Springel}, {Yoshida}, \&
  {White}}]{springel2001}
{Springel}, V., {Yoshida}, N., \& {White}, S.~D.~M. 2001, \na, 6, 79

\bibitem[{{Valluri} \& {Merritt}(2000)}]{valluri2000}
{Valluri}, M. \& {Merritt}, D. 2000, Advanced Series in Astrophysics and
  Cosmology, 10, 229

\bibitem[{{van den Bosch} \& {Ogiya}(2018)}]{vanderbosch2018b}
{van den Bosch}, F.~C. \& {Ogiya}, G. 2018, \mnras, 475, 4066

\bibitem[{van~den Bosch {et~al.}(2018)van~den Bosch, Ogiya, Hahn, \&
  Burkert}]{vanderbosch2018}
van~den Bosch, F.~C., Ogiya, G., Hahn, O., \& Burkert, A. 2018, \mnras, 474,
  3043

\bibitem[{{van Elteren} {et~al.}(2014){van Elteren}, {Pelupessy}, \& {Portegies
  Zwart}}]{amuse2}
{van Elteren}, A., {Pelupessy}, I., \& {Portegies Zwart}, S. 2014,
  Philosophical Transactions of the Royal Society of London Series A, 372,
  20130385

\bibitem[{{Walker} \& {Pe{\~n}arrubia}(2011)}]{walker2011}
{Walker}, M.~G. \& {Pe{\~n}arrubia}, J. 2011, \apj, 742, 20

\bibitem[{{Zhan}(2006)}]{zhan06}
{Zhan}, H. 2006, \apj, 639, 617

\end{thebibliography}

\end{document}